%% file: robu.tex
\documentclass[letterpaper]{article}
\usepackage[numbers]{natbib}
\usepackage{microtype}
\usepackage[hidelinks]{hyperref}
\usepackage{caption}
\usepackage{url}
\usepackage{tikz}
\usepackage{amsmath}
\usepackage{amssymb}
\usepackage{xcolor}
\usepackage{graphicx}
\usepackage{grffile}
\usepackage{setspace}
\usepackage{enumitem}
\usepackage{framed}
\usepackage{csquotes}
\usepackage{booktabs}
\usepackage{subfig}
\usepackage{svg}
\usepackage{wrapfig}
\usepackage{bashful}
\usepackage{pdftexcmds}
\usepackage{listofitems}
\usepackage[toc,page]{appendix}
\newcommand\crule[3][black]{\textcolor{#1}{\rule{#2}{#3}}}
\usetikzlibrary{positioning}
\usetikzlibrary{arrows}
\usetikzlibrary{automata}
\usepackage{pgfplots}
\IfFileExists{danielpar}{}{\pgfplotsset{compat=1.14}} 
\usetikzlibrary{positioning}

\title{Measuring Time with Minimal Clocks}
\author{Andrei D. Robu$^{1,*}$, Christoph Salge$^{1,2}$, Chrystopher L. Nehaniv$^{1,3}$, \and Daniel Polani$^{1}$ \\
\mbox{}\\
$^1$Adaptive Systems Research Group, School of Computer Science\\
University of Hertfordshire, Hatfield, UK  \\
$^2$Game Innovation Lab, Department of Computer Science and Engineering\\
New York University, New York, USA \\
$^3$Algebraic Intelligence \& Computation Lab, Faculty of Engineering,\\
University of Waterloo, Canada\\
$^*$Corresponding author. E-mail at a.robu@herts.ac.uk.}

\begin{document}

\maketitle

\begin{abstract}
  Being able to measure time, whether directly or indirectly, is a
  significant advantage for an organism. It allows for the timely
  reaction to regular or predicted events, reducing the pressure for fast
  processing of sensory input. Thus, clocks are ubiquitous in biology.
  In the present paper, we consider minimal abstract pure clocks in
  different configurations and investigate their characteristic
  dynamics. We are especially interested in optimally time-resolving
  clocks. Among these, we find fundamentally diametral clock
  characteristics, such as oscillatory behavior for purely local time
  measurement or decay-based clocks measuring time periods of a scale
  global to the problem. We include also sets of independent clocks
  (``clock bags''), sequential cascades of clocks and composite clocks
  with controlled dependency. Clock cascades show a ``condensation
  effect'' and the composite clock shows various regimes of markedly
  different dynamics.
\end{abstract} \hspace{10pt}

\providecommand{\keywords}[1]
{
  \small	
  \textbf{\textit{Keywords---}} #1
}
\keywords{timekeeping, information theory, clocks, oscillation, cycle, transfer entropy}

\pagebreak

\section{Introduction}
\begin{quotation}
  Time in itself, absolutely, does not exist; it is always relative to
  some observer or some object. Without a clock I say 'I do not know
  the time'.

  \mbox{}\hfill \textsc{John Fowles, \'{A}ristos}
\end{quotation}

In the present paper, we study how time measurement could
look in the most minimal cases which we believe relevant to
simple organisms.

\citet{klyubin2007representations} look at maximizing information flow
for a simple navigation task in minimal agents as proxy for other more
specific and/or complex achievements, such as homing. One striking
phenomenon that appears there is that when they consider information
processed by the agent, they find that, via a direct factorization
analysis, the internal processing of information by the agent can be
roughly decomposed\footnote{Other factorizations could be considered
  such as the multivariate information bottleneck
  \cite{slonim06:_multiv_infor_bottl} or decompositions based on
  unique information \cite{Bertschinger:2014iu} or complexity-based
  decomposition \cite{Ay2015}.} into an essentially
spatial and a temporal component. Specifically, the memory of the
informationally optimal minimal agent typically contains information
about the respective lifetime step of the agent without time ever
having been optimized for. In that work, the ability to measure time
emerged as a pure side effect.

Here we turn the measurement of time into our primary objective: we
consider measuring time as the most intrinsic, least environmentally
affected processual quantity, with the following assumptions: time has
a well-defined beginning; its tick is discrete, global and accessible
to the agent.

We will discuss the motivation behind some of these assumptions and
their implications in more detail in the following sections.
For now, it will suffice to say that
the first assumption is quite natural, as many biological processes
have a natural beginning, e.g.\ some trigger event, which might even
be just the beginning of the organism's morphogenetic process. The
discreteness of the tick constitutes a more serious conceptual
problem, as it poses the question concerning the origin of the natural
timescale. We will discuss this point later in more detail in
connection with external rhythms (or Central Pattern Generators,
CPGs). The final issue is the most subtle one, namely how an external
tick becomes accessible to the organism (assuming it is not a
self-generated CPG oscillation). This major issue does not fall into
the scope of the present paper and will be considered separately in
future study.

\section{Motivation}
\label{sec:motivation}

\subsection{Why Do We Study Clocks in Minimal Systems?}
As mentioned above, \citet{klyubin2007representations} showed that
(imperfect) time measurement can emerge as consequence of an agent
optimally extracting information about its environment. This
information that the agent acquires in its memory is partially
determined by the environment and partially intrinsical (via the
discrete time tick of the model). That work, however, shows an
incomplete (yet more plausible) case of time
acquisition.

We, however, believe that it is studying the extreme cases of a
phenomenon that is essential to understand better the constraints and
conditions that are operating on a potential artificial life (ALife)
system and what fundamental effects they elicit. In this sense,
studying pure time measurement is, in a way, studying an extreme case,
namely the ``most intrinsically acquirable'' information,
requiring only a tick and an initial state to compare with and no
further environmental structure/input. This extreme minimal setting is
what we set out to study here --- to ask the question what time
measurement could possibly look like in an informationally limited
system.

Is this relevant for survival? We know from
\citet{klyubin2007representations} that some time-keeping at least
emerges as a side effect for a more-or-less navigational task --- and
a hypothesis hovering in the background is that, once discovered, good
timekeeping might then be used by a species for exaptation to novel
niches where it can complement other information which
otherwise were not of sufficient quality.

While the argument for this is too long to be made in full in the
present paper, we will sketch it briefly here: an agent needs to
acquire a certain amount of \emph{relevant Shannon information} from
the environment to carry out a given task at a given performance level
\citep{salge2010relevant, polani06:_relev_infor_optim_persis,tishby11:_infor_theor_of_decis_and_action,Still2012}.
However, in some quite generic ALife-type scenarios it can occur that
a sometimes significantly larger amount of Shannon information needs
to be acquired by an agent from the environment than corresponds to
the actually relevant information. This can be seen as the
information-theoretic analogy to thermal efficiency from physics; or,
in the information-theoretic case, its complement, namely the
informational inefficiency and closely related to the ratio between
excess entropy and statistical complexity
\citep{shalizi01:_causal_archit_compl_self_organ}. As such, this is
quite a generic phenomenon and its occurrence is not the exception,
but the rule.

It turns out that this excess or ``piggyback'' 
information, while not
relevant to the original goal, can sometimes be repurposed for another
goal \citep{dijk12:_infor_drives_sensor_evolut}. It was hypothesized earlier that such a phenomenon might be
pervasive in evolution and in fact offers explanations for some
interesting evolutionary phenomena such as the ability for exaptation or
very rapid refinement and specialization of sensors.

Other, subtle reasons why time measurement is essential to understand
the nature of cognitive control by an agent is the computational
complexity required to keep track of time \citep{2018arXiv180604103H};
these require further studies and are, partly, still open questions
themselves. For now, we choose to concentrate on as much simplicity
and minimality of the model as we can muster.

\subsection{The Discrete, External Tick}

The clocks in this paper are modeled with a discrete and finite (and very small) state space with a discrete axis of time.
Pairing a discrete space world with a discrete axis of time means that
all clocks have regular ticks. In a
model where multiple clocks are simulated together, this also implies
a synchronized global tick for all subclocks in the system.

A model with an implied regular tick given to all clocks is a
considerable assumption. By making this assumption, we are choosing to
not study how a synchronized and regular ticking mechanism could
emerge and be sustained in a world with continuous time. Instead, we
focus entirely on which configurations of minimal clocks we can
possibly find once a regular, overall synchronized tick is assumed to
exist. Notably, we completely ignore the route towards how such a
tick would come about in the first place. We note that emergence and
mechanisms of CPGs (see also the remarks in the section on continuous
time further below) are thoroughly studied topics. In our context, the
analogous question would consist of the information-theoretic
characterization of the process of how synchronization, adaptation and
reliability would emerge in minimal oscillators (ultimately with
biological plausibility in mind). However, this is a topic that
requires a substantial separate study and will be studied in future
work.

\subsection{Sensing Time vs. Measuring It}

Before we proceed, we wish to discuss why we do not include into our
considerations the simpler possibility of utilizing external cyclic
behavior as a possible paradigm for time measurement.

Generally, having an external ``timer'' helping an agent to
synchronize with its surroundings is a useful approach and probably
altogether unavoidable once one moves into continuous time. However,
in this paper, we restrict ourselves to considering the extreme case
of an almost solipsistic system where the only environmental influence
is the defined starting state and the global discrete tick.
Emphatically, this model generalizes and morphs naturally into any
system with an explicit external clock. The latter, however,
constitutes an informationally more intricate case: while it makes
time measurement easier for the organism, it renders the analysis of
the ``pure'' time measurement considerably more difficult and will
have to be left to future study.

\subsection{Continuous Memory}

The study of clocks could also be imagined to be conducted in a
continuous-space world (i.e.\ the state space of the clocks is
considered to be continuous). We choose to model clocks with discrete
state spaces because, in a model with continuous state spaces (and
lacking any other constraints such as imposed noise), clocks optimized
for time resolution would end up being unbounded in their memory
capacity (and also complexity). There, an efficient optimizer could
potentially find arbitrarily complex clocks when maximizing
information about time.

One possibility to limit this complexity while using continuous state
spaces would be to introduce noise which would naturally limit the
achievable resolution. However, this would imply the introduction of
additional assumptions. Since our focus is primarily the study of
minimalistic clocks, we choose arguably the most direct way to
restrict the complexity of the clocks, by limiting their state
spaces to be finite (and small).

\subsection{Continuous Time}

Central Pattern Generators (CPGs) are biological neural circuits that produce oscillatory
signals. \citet{marder2001central} define them as:

\blockquote{
    [...] neuronal circuits that when activated can produce rhythmic motor patterns such as walking, breathing, flying, and swimming in the absence of sensory or descending inputs that carry specific timing information. 
}

CPGs are common in biology. They implement ``soft'' clock
ticks in the form of oscillations, in continuous time and having a continuous
state space. They reduce the burden on other units in a
biological system of generating ticks or defining phases.

CPGs are examples in nature of how the task of generating regular ticks can
be a specialized one, segmented from other processes.
And, similar to the way CPGs commonly operate in biological systems
to provide ticking to other mechanisms, we hypothesize that in general,
clocks can also be seen as being composed of a tick producer 
and their (possibly probabilistic) tick counter. Of the two parts,
we focus our study in this paper on what are essentially probabilistic tick counters,
instead of studying the tick generating mechanisms. 
By modeling the axis of time as made up of discrete moments, the
simulated clocks implicitly receive a regular clock tick. Their stochastic behavior
is applied once at every time transition.

Since in the present paper we consider probabilistic clocks, these
would permit meaningful models for continuous time even with a finite
number of states. Now, in the present paper we are explicitly
interested in the edge cases of alternators (discrete oscillators) and
the possible emergence of counters on top of these.

However, \citet{Owen_2019} find that displaying an oscillatory behavior
with finite states in continuous time at equilibrium is impossible
without a hidden state. In their paper, they show that under the
conditions described therein, to implement oscillations in continuous
time, one needs a state where the distribution can move into while
between transitions. Using a Markov chain as model, that Markov chain
requires at least a triad of states in a discrete system in continuous
time. While we do not reiterate the precise (and intricate) argument behind
the work by \citet{Owen_2019}, one intuition behind this is that in
continuous time one essentially cannot move probability masses from
one state to another arbitrarily fast and they have to be moved
continuously. This necessitates at least a third state to store it.
The analogy in a continuous state space would be that a Markovian
implementation of an oscillator would require a minimum of two
dimensions to realize an oscillation (state and rate of change,
corresponding to the second order differential equation for
oscillatory dynamics).

Since in the present paper we do not wish to model either hidden or
external states, but wish to encompass the whole system, we will again
postpone this possible generalization of our scenario to a future
study and limit ourselves to modeling discrete-time dynamics of
discrete state systems which will already exhibit a richly structured
set of phenomena.

\subsection{General Comments on Time Measurement}

More specifically, in the context of biologically plausible models, we
are interested in how a Markovian agent can keep track of the flow of
time under strong limitations on memory capacity. Going forward with
the earlier assumption of the pre-existence of ticks, measuring time
becomes effectively equivalent to counting. We will find
that, to make the best out of limited resources, in some circumstances
we will need to count probabilistically. This is what we will study
in the present paper.

In its conceptually simplest incarnation, the measurement of time
would consist essentially of two components: having a reliable
generator of periodic behavior (which we pre-assume in
the present context); and being able to count the periods.
To measure larger time intervals precisely, one needs
full-fledged counters, which, in turn, require a comparatively complex
logical make-up (and more complexity in temporal structure --- such as
with multiple distinct event types, multiple kinds of 'clock ticks' or
'events' --- opens the way to the general structure of discrete
dynamical systems and their algebraic understanding via semigroup
theory, the general study of models of time --- see
Sec.~\ref{sec:AlgTime}), while not impossible in principle, one would
not expect this to appear generically in biologically relevant
scenarios and most certainly not in very simple organisms.

Rather, one would typically expect to find less precise, but simpler
and more robust solutions. The present paper investigates how the most
minimal of such models could look and which characteristic properties
we expect them to exhibit. In our discussion we include both explicit
clocks characterized by a distinct apparatus which clearly have a
time-measurement function such as the suprachiasmatic nucleus in the
mammalian hypothalamus involved in the control of circadian rhythms
\citep{klein1991suprachiasmatic}, but also implicit clocks (which
exhibit some aspects of clock-like behavior, such as being correlated
to time, but without a dedicated mechanism) like the feeding-hunger
cycle of an animal. We do not differentiate these classes a~priori
since our formalism does not discriminate between them. Instead, our
study will concentrate on systems which, given their constraints, are
maximally able to measure time, not considering any other tasks ---
thus we study clocks as ``pure'' as they can be under the
circumstances given. 
We expressly are not studying how an agent could
infer time from correlations in the environment (using an --- external
--- sundial for example), but only in how time can be tracked
intrinsically. For such a study of the evolvability and robustness of internal time-keeping mechanisms in gene regulatory networks in the context of (possibly intermittent and noisy) external periodic variation, see e.g.\ \citep{KnabeAL2008}.
 In the same vein as considering Artificial Life as
about understanding life-as-we-know-it vs.\ life-as-it-could-be
\citep{langton89:_artif_life}, and in view of how life pervasively
makes use of clocks, the present paper studies possible clocks
themselves, or, in analogy to Chris Langton's research program:
``time-as-it-could-be-measured''.

\section{Temporal Dynamics: The Algebra of Time}\label{sec:AlgTime}

We begin with some general comments on the structure of time. Whether in classical or relativistic physics, or
in more general models (e.g., ancient Indian notions of cyclical time,
or in modern automata networks), there is a commonality to notions and
models of time: 
From the perspective of a single organism or agent,
events in time satisfy a grammatical constraint attributed to Aristotle in \citep{rhodes2009applications}:
If $\alpha$, $\beta$ and $\gamma$ are each sequences of events, then:
if $\gamma$ follows $\beta$ in an agent's experience, and, prior to both, $\alpha$ occurs,
that is exactly the same as when $\beta$ follows $\alpha$, and $\gamma$ occurs after both, i.e.\ 
\[ \alpha(\beta \gamma) =  (\alpha \beta) \gamma .\]  

In short,  sequences of events in time from the perspective of an  individual agent
satisfy the  {\em associative law}.  A structure with a `multiplication' operation satisfying this law is thus a model of time, called a {\em semigroup} in mathematics.
Here, concatenation is silently used to denote the binary operation of following one sequence of events by another, which is the associative multiplication operation of the semigroup of event sequences.\footnote{Equivalently, we can identify events and sequences of events with operators transforming the agent's world, in which case the associative operation is the composition of operators.}

Specifically for time which has only a single kind of clock tick $t$,
let $t^n$ denote the occurrence of $n$ clock ticks one after another.
One can distinguish classes of  time which exhibit only a
single type of ticks (or event) $t$
as follows: determine the first $k>0$ for which there is an $\ell>0$
such that $t^k=t^{k+\ell}$, and whether such a $k$ exists or not. Choosing
$\ell$ minimal, time cycles every $\ell$ ticks after an initial
transient $t, t^2, \ldots, t^{k-1}$. There are then four qualitatively
different kinds of single clock tick-generated time
\citep{nehaniv1993algebra}: (a) cyclical time ($k=1$), where changes
repeat in cycles of exactly $\ell$ steps, (b) purely transient time ($\ell=1$) --- with
no change happening after $k$ ticks, (c) a transient followed by a
cycle $(k>1,\ell>1$), where after $k$ initial transient steps time becomes an $\ell$-step cycle, and, finally, (d) infinite non-repeating discrete time
indexed by positive integers, in the remaining case when no positive $k$  and $\ell$ exist with $t^k=t^{k+\ell}$.  
In this classification, $\ell$  is the length of the attractor cycle that time eventually must enter, if $k$ exists; and then $k-1$ is the length of the transient before the attractor is reached.
Case (d) can be viewed as having an infinite transient, $k=\infty$, with no attractor, $\ell=0$. 

Note that all these models of time based on a single type of clock tick are necessarily
{\em commutative}: in other words, here the order of event sequences
does not matter, i.e., $t^a$ followed by $t^b$ is the same as $t^b$
followed by $t^a$, since both are comprised of exactly $a+b$ clock
ticks $t$. (Every event sequence $\alpha$ is some number of ticks, i.e., $\alpha=t^a$ for some $a\geq 1$, and any other sequence $\beta=t^b$ (for some $b\geq 1$) is too. Thus it follows $\alpha \beta = \beta \alpha$.) 
This is a special property of semigroups generated by a single event type, which are classified as above into four types. Non-commutativity becomes possible as soon as there is more than one type of event or clock-tick, e.g., when the agent can choose between two or more actions, or undergo at least two different kinds of events, whose order may not be deterministic. 

This type of modeling of time as a semigroup is common practice in
dynamical systems theory. Note that it is in general not a {\em
  group}, which would be a model of time where every sequence of
events must be reversible; in particular, there
may be an `earliest time' or  multiple `Garden-of-Eden'  times beyond which a process
may not be reversed (whenever there is a non-trivial transient, i.e., $k>1$); 
furthermore, the process need not be reversible
at all. Note that among the above models, cyclic time with
$k=1$ (case (a)), forms a group where enough repetitions of $t$ reverse any
sequence, and the infinite counter (case (d)), embeds in a group: time indexed by
the integers $..t^{-2}, t^{-1}, t^0, t, t^2,..$. 
When $\ell\geq 0$, the cycle  $t^k, ..., t^{k+\ell-1}$ comprises a
group substructure of the model of time, i.e., it implies a {\em local pool of reversibility}. 
Note that with these models of time, one can never reverse events
suitably to re-enter the transient part of the
dynamics.\footnote{There is a subtle point here for the infinite case
  (d):   if the clock tick $t$ is realized as an operator on 
  agent's world (mapping states to states), the embedding of positive natural numbers $\mathbb{N}^+$   (the
  infinite transient) into all integers $\mathbb{Z}$ extends to an embedding of
  operators if and only if the operator $\varphi(t)$ corresponding to $t$ is itself
  invertible (with its inverse associated to $t^{-1}$, and so that
  $t^0$ gives the identity operator).  In the latter case, once the
  embedding is done, one could assert of the resulting system that
  $k=1$ (`there is no [longer any]
   earliest time') and $\ell =\infty$ (`the apparent transient has become part of
  an infinite, reversible cycle'). That is,  the embedding $\mathbb{N}^+\hookrightarrow\mathbb{Z}$  induces
  an corresponding embedding for the powers of the operator $\varphi(t)$, since negative powers of
  the operator $\varphi(t)$ are well-defined. (This fails when $\varphi(t)$ is not invertible.)
    See \cite{NehanivSYDE710} for full details relating events and operators in models of time.}

The study of (possibly general) structures satisfying these laws becomes
the study of models of time. This viewpoint has deep connections with
the theory of discrete or continuous dynamical systems
\citep{rhodes2009applications,NehanivSYDE710}. In the finite discrete deterministic
case it leads to the Krohn-Rhodes theory, a branch of mathematics
(algebraic automata theory) where discrete dynamical systems can be
decomposed using (non-unique) iterative coarse-graining into a cascade
of irreducible components. The composite clocks discussed later
constitute a special case of such a decomposition.

With more event types than time ticks (e.g.\ consider an agent which
has the choice of different actions and thus ``futures''), one has to
generalize further. Multi-event semigroups can be interpreted as
agents that can have multiple alternative timelines. Here, in
particular, the commutative law no longer holds in general: Washing
ones hands before eating, or putting a glass on the table and pouring
water has not the same outcome as when the events occur in the reverse
order, but has a markedly different outcome from the other way round.
Thus, in this case, $\alpha \beta$ is not necessarily the same as
$\beta \alpha$. In this case, processes in time are not commutative;
in these cases a more complete picture of semigroup theory needs to be
invoked to fully describe the scenario.
  
In the remaining sections of the paper, we will not refer to different
event types, but instead consider nondeterministic representations of
time which emerge through informational optimality criteria. This
causes much more complicated and interesting dynamics to arise that
partly (but only partly) mirrors some of the aspects of the semigroup
decomposition (see Sec.~\ref{sec:composite-clock}). It turns out that,
even just considering the simple semigroup of time, the probabilistic
setting under informational optimality gives rise to a rich and
complex landscape extending and generalizing the discrete semigroup
model into the probabilistic continuum.

We would note that the paper is at this stage not going to consider
the case of multiple time observers, but only a single observer with a
coherent tick. It will be insightful to consider the generalization to
multiple observers. Even without considering relativistic scenarios,
having non-synchronized observers (such as multiple sub-organelles in
an organism, or, on a larger scale, the existence of local times before
the introduction of standardized time zones or precise clocks due to
the necessity to fit together train schedules) complicates the
situation considerably.

\section{The Cost of Measuring Time}
Recent work by \citet{barato16:_cost_precis_brown_clock} highlighted
interest in the problem of time measurement by asking if clocks must
pay a thermodynamic cost to run \citep{2018arXiv180604103H}. Their
stance is grounded in fundamental trade-offs of physics which cannot
be subverted. But, at the level of organisms which are far remote from
those physical trade-offs, those physical limitations, conservation
laws and constraints do not apply in a straightforward manner.
Concretely, we work in a near-macroscopic, classical (non-quantum)
Markovian universe without presuming the additional structure of
microphysics (including microreversibility or Hamiltonian dynamics).
In particular, we cannot assume an obvious generalization of the
physical concept of energy to a fully general Markovian system. Thus,
it is not obvious how to quantify computation cost in terms completely
analogous to thermodynamics. The only concepts that carry over are of
entropic and information-theoretic nature. Specifically, Shannon
information has been shown to be a highly generalizable measure of
information processing cost \citep{polani2009information}; a universal
measure that can be used to compare systems of a different nature, and
does not presuppose any structure on the state space of events; it
permits to ignore labels and only considers the statistical
coincidences.

Henceforth, all costs will therefore be expressed in the language of
information theory; they will refer essentially to information storage and
communication costs and constraints.

\subsection{Small State Spaces}
In the beginning of this paper we focus on the most minimal clocks
(and later proceed to slightly more complex arrangements) for a number
of reasons. For one, this will make it easier to explore the full
solution space. The second reason is significantly more subtle, and we
will only be able to sketch it here: essentially, it is not clear what
measure of complexity to utilize for the cost of running a larger
single-component counter and/or the complexity of running the
transition itself; natural candidates for such costs might be
predictive information \citep{bialek01:_predic}, statistical
complexity \citep{crutchfield89:_infer} or forecasting complexity
\citep{grassberger1986toward}. However, the possible candidates are
not limited to these three measures, and many other plausible
information-theoretic alternatives can be conceived\footnote{We will
  investigate this question in a separate paper, but for a discussion
  of the issue of computational complexity in a Shannon context, see
  e.g.\ \cite{2018arXiv180604103H}.}. In absence of a canonical
measure for the complexity of a single-component clock operation, here
we limit ourselves to investigate the most minimal clocks possible,
namely a 2-state (1-bit) clock and we will here not further concern
ourselves with taking into account the informational cost of actually
running this clock.

\subsection{Information Flow Between Modules}
We said above that we prefer to start out with already minimal clocks
by default to avoid dealing with the question of how complex the
operation of a clock is. But what if an agent needs a larger clock? In
this case, we will  build it out of smaller clocks. Again, we avoid
the question how to cost this compositional complexity
\citep{2018arXiv180604103H}.

As we will state more quantitatively below (Sec.~\ref{sec:composite-clock-experiment}), such a
compound clock will perform better if its components ``cooperate''.
For them to be able to cooperate, they must be able to exchange
information. Therefore any limitation in this clock's internal
information flow will reduce the performance of the clock.

The motivation for limiting the information flow is that such a flow
will in general be costly in biology, even outside of the earlier
mentioned thermodynamic considerations; information processing in
biology is expensive per se (even far away from the Landauer limit)
\citep{laughlin01:_energ}. This will limit how many components can
communicate with each other and at which bandwidth.

All in all, in our studies below, when looking for candidates for
clocks, we will consider Markov chains with a small number of states
and when we move on to larger clocks we prefer to build them out of
minimal clocks and use the information flow between the components as
cost to consider.

\section{Other Relevant Work}
We mentioned earlier the work by \citet{klyubin2007representations}
who consider the maximization of information flows in a very simple
agent/environment system. The resulting agent controllers generate a
rich set of behaviors, with a side effect that the controllers'
dynamics causes the agents' internal states to partially encode
location, but also partially time information. More precisely, when
one inspects the agents' memory, it turns out that it provides partial
information about the point in time from the beginning of the
experiment. Some information about time and space can be extracted
from the agent memory separately through a factorization process.
Note, however, that, in that experiment, the joint encoding of
spatiotemporal state in the agents' memory is merely a side effect of
the homing information optimization task and not directly optimized
for.

The importance of measuring specifically time (as opposed to having
this measurement emerge via side effect) can be seen in other
scenarios, for instance in 13- or 17-year cicadas
\citep{karban00:_how,sota13:_indep}. Also, closer to the level of
fundamental physical limits, \citet{chen2010clocks} find relations
that can be obtained by applying Fisher information to the problem of
measuring time with quantum clocks and discuss the problem of clock
synchronization in the quantum realm.

On the issue of considering global ticks, we note that
\citet{karmarkar2007timing} mention mounting evidence against the view
that the brain keeps track of time by counting ticks and instead study
the behavior of simulated SDNs (state-dependent neural networks),
showing how neural networks can keep track of time implicitly in their
states without counting \citep[note the parallelism to the indirect
identification of time
in][]{klyubin2007representations}. We also note
\cite{van2012dynamic} which discusses not only the possible mechanisms
of time perception in the brain but also the difficulty of validating
them.

Since we hypothesize that our abstract considerations find functional
correspondences in nature, the simplest examples would be expected to
be found in microorganisms. We would predict that, if our general
hypothesis is correct, the characteristics of our results for minimal
clocks will be reflected in very simple organisms. We will preempt
here one result detailed in the results section. Under constraints on
the memory available to the clock, only two types of clocks will be
found --- local, short-term clocks  measuring the time within a cycle
(essentially its phase); and, long-term clocks which distinguish
large-scale phases within an overall ``lifetime'' interval of
interest. Thus, we get a very clear dichotomy between local time
measurement and global time measurement. We will call the first type
cyclic clocks or oscillators, and the second type ``drop clocks''
(essentially one-off decay-type time measurements). Indeed, it turns
out that bacteria show examples of both cyclic and drop clocks. Examples
are in the following paragraph.

\citet{hut2011evolution} discuss the reasons why day and night require
different behaviors, these forming the selective pressures of
evolution for the circadian rhythm of bacteria --- the periodic cycle
of activity in organisms that allows them to adapt to the time of day.
This is one central example demonstrating the importance of
time-keeping to organisms. They note that the circadian rhythm of
cyanobacteria has been reproduced and studied in a test tube.
Therefore this is an example of a relatively simple, clearly
understood explicit cyclic or \emph{oscillator} (see below) clock.
\citet{nutsch2003signal} study prokaryotic taxis for halobacteria, more specifically signal transduction pathway starting from sensing light and responsible for controlling the switching of the flagellum. This pathway implements a one-off drop clock\footnote{Which is reset only by an external trigger; some mechanisms can be considered drop clocks  triggered at conception and never reset until death, such as telomeres.}.

They note that while the molecules (the chemistry) involved in this pathway are well
studied, the way these components behave together dynamically is not
well understood and only speculated on. Building upon an earlier model
by \citet{marwan1987signal}, they suggest dynamical models to fit
experimental findings. These models are synthetic and not based on
first principles. Thus, the question about the actual structure and
size of the fundamental clocks of  bacteria remains currently
unanswered.  

Remarkably, measuring time can also help bacteria in spatial tasks.
Some bacteria use differences along the length of their body to
measure gradients \citep{oliveira2016single}, however, others instead
measure time differences between intensities while moving
\citep{nutsch2003signal}. This is demonstrating how measurement of
space can be converted into measurement of time, emphasizing once more
the intimate relation between time and space in even the minimal
navigational capabilities of bacteria. 

\section{Models of Clocks}
\emph{In the real world, time is always encoded in physical systems; in a broad sense, any evolving system with nontrivial dynamics can be regarded as a clock. To read the time, we must perform measurements and estimate the parameter that encodes the time. How precisely we can estimate this parameter depends on the distinguishability of the involved states, and this distinguishability characterizes the quality of a clock.} ---~\citet{chen2010clocks}

Because our clocks are modeled as Markov chains, they are completely
described via their current state and their probabilistic dynamics.
We consider only discrete state spaces and also advance in time only
in the discrete steps given by the global tick. We reiterate that this
means that the clocks receive time ticks for free and that they need
not be concerned with the challenge of getting accurate ticks. Apart
from these global ticks, the clocks do not receive any information
from the environment. We furthermore consider first only the most
minimal clock designs. As we proceed, we will consider more complex
(composite) clocks.

\subsection{Notation}
\label{sec:notation}

Random variables are written as
capital, their values as lowercase letters. The probability of a
random variable $X$ to adopt value $x$ will be written as $P(X=x)$ or,
where not ambiguous, as $p(x)$ by abuse of notation. The state of a
clock which can take on values $u$ and $d$ (``up'' and ``down'') will
be denoted by random variables $S$ (possibly subscripted by time $t$,
because in general the probability distribution over the states of the
clock changes in time). To model the uncertainty that the agent has
about the current time, we treat true time (which the clocks attempt
to measure), similarly to \cite{klyubin2007representations}, also as
random variable $T$ which a~priori assumes all possible (integer) time
values between $t=0$ and $t=T_{\text{max}}$ with equal probability.

Typical quantities discussed would be, for instance,
$P(S=u|T=t) \equiv p(u|t)$, the probability that the clock state is
$u$ (``up'') at time $t$. Or else, $P(T=t|S=d) \equiv p(t|d)$ would be
the probability that the time is $t$, given that the current state of
$S$ is $d$, etc. When we optimize clocks, we quantify our criterium of
performance as mutual information $I(S;T)$. This mutual
information tells how much information a previously uninformed agent
would receive about time after looking at its clock (averaged over
probabilities of the possible states observed in the clock, ``up'' or
``down''). The quantity $I(S;T)$ is directly related to the
probability of guessing time correctly given one observation of the
current state of the clock. With this notation, we are ready to define
our clock model. In all our experiments, the clocks will be
initialized at time $t=0$ in a fixed known state, specifically $u$,
i.e.\ $P(S_0 = u) \equiv P(S=u|T=0) := 1$ and they will advance by one
time step per tick according to their respective dynamics.

\subsection{Oscillator}
\label{alternator-clocks}
We first define the \emph{oscillator} as the 2-state Markov chain with a symmetric
probability to change state.

\begin{figure}[htbp]
\centering
\input{figs/robu.alternator-state-diagram.tex}
\caption{The oscillator clock. It switches one state to the other with probability $r$.}
\end{figure}
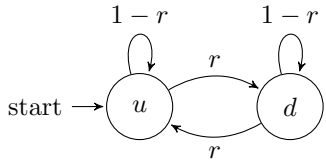

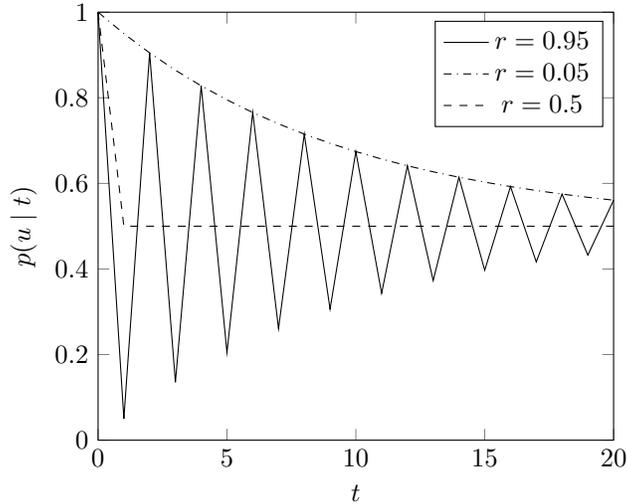
\begin{figure}[htbp]
\centering
\input{figs/robu.alternator-time-behaviour.tex}
\caption{The oscillator clock for different probabilities of switching state.}
\label{fig:alternator-time-behaviour}
\end{figure}

The plot of ${p(u \mid t)}$ in
Figure~\ref{fig:alternator-time-behaviour}, shows the 
distinction between different regimes in which
the clock operates, depending on $r$. We notice the analogy with
Damped Harmonic Oscillation.
\begin{description}
    \item[ Stuck ]
      For ${ r = 0 }$, the clock is stuck in state $u$ (not shown in diagram).
    \item[ Overdamped ]
      For ${ 0 < r < 0.5}$, the probability distribution behaves like
      an overdamped oscillator.  Note that, in this regime, the
      probability distribution of the oscillator clock has the same
      envelope as the drop clock which we discuss below. 
    \item[ Critically Damped ] For ${ r = 0.5 }$, the clock acts
      analogous to a critically damped oscillator: it reaches the
      equilibrium in the shortest possible amount of time --- one time
      step in our discrete time case.
    \item[ Underdamped ] For ${ 0.5 < r < 1 }$, the probability
      distribution behaves like an underdamped oscillator. This can be
      interpreted as the clock being initially ``synchronized'' with
      time like a clock, but that the synchronization gets lost as the
      clock's state changes in time, until the correlation between $t$
      and $s$ disappears.
    \item[ Undamped ]
      For ${ r = 1 }$, the clock state alternates between $u$ and $d$. Not shown in figure.
\end{description}
These behaviors are shown for the symmetric oscillator, but the
asymmetric one (where the transition probabilities for $u \rightarrow
d$ and $u \leftarrow d$ differ) has the same qualitative behavior,
just with a different equilibrium point. All 2-state Markov chains
belong to one of these 5 classes. In other words, the class of 2-state
clocks can exhibit only this small number of different behaviors. We
will refer to this insight in the results section again. 

\subsection{Drop Clock}
\label{drop-clock}
Consider the thought experiment of an insect that leaves its nest to
forage or explore. Even if it does not find anything, the insect
should still initiate a return to its nest at some point in time,
otherwise it risks getting lost. Such an insect
would profit from  a clock  telling it if ``it's been awhile'' since
it left its nest.

An oscillator is not well suited for this type of task; it is able to
measure a local phase (an odd or even step), but it does not provide
much ``larger timescale'' information, unless it is overdamped.
Instead of the latter, it turns out that a \emph{drop clock} is a more
natural model for this task.

More complicated models will exist in nature. More complex clocks,
such as clocks with time-dependent transition laws, however, require
more memory under the Markovian constraint. Another example is that of
gene regulatory networks which extends beyond discrete time into
continuous time. Here, however, our question strictly focuses on the
simplest possible ones.

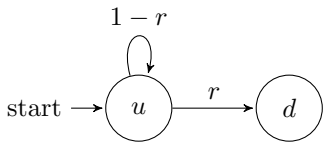
\begin{figure}[htbp]
    \centering
\input{figs/robu.drop-clock-state-diagram.tex}
    \caption{The drop clock, a Markov chain with a probability $r$ to permanently transition to state $d$ and remain there.}
\end{figure}

The drop clock starts (as all our clocks) in a well defined state,
namely $u$, i.e.\ with $P(S=u | T = 0) := 1.$ After each time step,
there is a probability $r$ that the clock ``decays'' (transitions from
state $u$ to state $d$) and the counterprobability $1-r$ that it does
not. Once the clock has decayed, it remains in state $d$ forever. The
behavior that this rule generates is an exponential decay $(1 - r)^t$
in time. An agent using a drop clock infers probable time only from
the state of this clock without having any other  knowledge of
what the time is.

\section{Experiments with 1-bit Clocks}
Our first experiments are dedicated to the 1-bit (i.e.\ 2-state) clocks. As discussed earlier, there are only few distinct classes of such clocks. Most notable are the 2-state oscillator and the drop clock.
The oscillator offers the maximum of 1~bit of information about time, namely whether one is in an odd or even time step. However, this information is purely local and cannot distinguish whether one is in an early or a later section of a run.
However, even with only 1~bit of state, the drop clock can provide
this distinction, albeit quite imperfectly, typically at significantly
less than 1~bit resolution. For it to provide best results, the
probabilistic decay rate of the drop clock (unlike the oscillator)
must be attuned to the length of the total time interval of interest;
this rate will be in general acquired by evolution or some learning
mechanism --- however, here, we will directly obtain it by
optimization of informational costs. This gives us a handle on what
the most effective possible drop clock could possibly be.

We study this next. Note that our axis of time is almost featureless
except for two features: length of time and the grain. The oscillator
matches the grain (local information) and the drop clock matches the
length (global information). As discussed earlier, we do not impose
any other features on the axis of time (such as months or
seasons, which constitute purely external drivers) because we only
study pure time, based on the fundamental tick only. For a meaningful
link to external events, as for a consistent treatment of continuous
time, one would need to adopt an approach more closely resembling the
study of \citet{klyubin2007representations}.

\subsection{Measuring Large Time Scales}
\begin{figure}[htb]
\centering
\input{figs/robu.drop-clock-phase-transition.tex}
\caption{The optimal drop probability $r$ for different timescales. Note that the curve for the drop clock has a discontinuity at ${|T| \approx 15}$.}
\label{fig:drop-clock-phase-transition}
\end{figure}
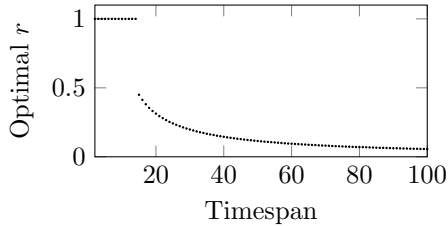

We now investigate time measurement by drop clocks at different
timescales. We tune the clock by finding the drop probability $r$
maximizing $I(S; T)$ for the particular timescale of the experiment.
Optimizing this with a parameter sweep (of values $0 \leq r \leq 1$
at increments of $0.01$) gives
Fig.~\ref{fig:drop-clock-phase-transition}. 

The first result is that the decay rate $r$ that best resolves time
with a drop clock clearly depends on the time interval. Such a clock,
therefore, must be adapted to the particular time interval to resolve.
Strikingly, we find two regimes of solutions, namely one with one
fixed decay rate (up to $T \approx 15$), and then a time
interval-dependent decay rate.

\begin{figure}[htb]
\centering
\input{figs/robu.drop-clock-2d-plot.tex}
\caption{Time information for different drop probabilities and time spans.
  Taking a slice in this plot at the dashed orange line creates the next plot (Fig~\ref{fig:two-maxima-drop-clock}). The yellow dots show the best decay parameter for that time span (essentially the same data in Fig.~\ref{fig:drop-clock-phase-transition}). The yellow contour lines are drawn for additional clarity.}
\label{fig:drop-clock-2d-plot}
\end{figure}
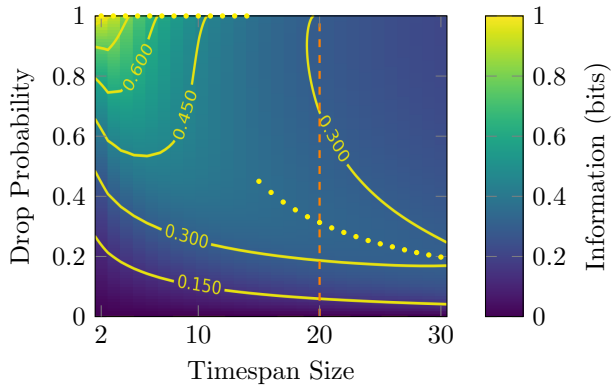

A closer inspection of Fig.~\ref{fig:drop-clock-2d-plot} shows a relatively complex landscape where a global maximum
of time information at the maximal decay rate $r=1$ is superseded at
larger times by maxima at lower decay rates.

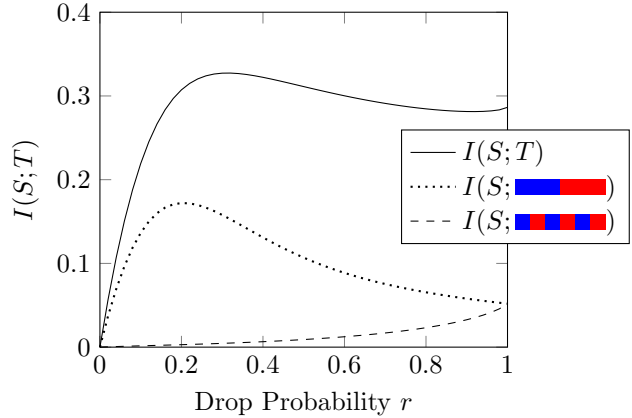
\begin{figure}[htb]
\centering
\input{figs/robu.two-maxima-drop-clock.tex}
\caption{Time information for different drop probabilities at ${T = 20}$. Apart from total time information, it shows information of clock state $S$ about being in an odd vs.\ even time step or  early vs.\ late phase of the measured period.}
\label{fig:two-maxima-drop-clock}
\end{figure}

Figure~\ref{fig:two-maxima-drop-clock} shows (solid line) a vertical
slice from
Figure~\ref{fig:drop-clock-2d-plot} (dashed orange line) at ${T = 20}$. This information
curve does not have a unique maximum, but an inflection. This
inflection stems from two separate contributions.

We can identify the two-fold origin of the two local maxima by partitioning
time into different features: one way to look at time is to distinguish an
``earlier'' from a ``later'' part and the other way is to distinguish
``odd'' vs ``even'' times. Our interpretation is that the different
maxima result from information the clock has about different
partitionings of time. The maximum around $r \approx 0.3$ appears to
come from the global picture the clock has (the ``earlier'' vs
``later'' partitioning \input{figs/robu.lr-variable.tex}\unskip) while the
second maximum at $r = 1$ comes from the ``odd'' vs ``even''
partitioning (\input{figs/robu.alt-variable.tex}\unskip).

Smaller (slower) decay rates prove better at resolving global timing
(early vs.\ late), and, while the drop clock generally resolves
odd/even times only weakly, it still achieves this level resolution
best for a hard decay rate for $r$, i.e.\ for decaying right away at
the beginning of the interval. This explains the phase transition in
Fig.~\ref{fig:drop-clock-phase-transition} between the regime of short
time spans to that of large time spans. This transition occurs because
of the inflection in the information curve of the sharply initialized
drop clock, when one maximum dips down below the other: different
parts of the curve derive from knowing different aspects about the
current time and because in some regimes one dominates the other.

We note that these different regimes make the drop clock hard to
optimize --- short time spans require a different strategy than long
ones: not only must the clock be attuned to a particular time scale to
best measure global time, but also does the information curve of the
clock show two maxima where one overtakes the other. We note that this
transition from one maximum to the other corresponds to a first order
phase transition, since they keep a finite distance from each other
when the transition occurs.
 
\subsection{Bag of Clocks}
\label{sec:bag-of-clocks}
Having studied the 2-state clocks, we now design a larger experiment.
We would like to keep the individual clocks simple, but be able to
measure time more accurately. For this purpose, we consider ``bags''
of independent non-communicating clocks to measure time. Notice that
even though the clocks in the bag cannot communicate with each other,
the total amount of information that the bag of clocks holds about
time will in general still be higher than that of the individual
clock. This is because each clock can hold information about a
different part of the axis of time.

We start the experiment with an empty bag. The collection is then
built up incrementally, one clock at a time. The clock that is
currently being added is optimized and after it is optimized so that
the current bag maximizes time information, the latest added clock is
frozen and added to the collection. We use this incremental process to
capture the intuition that, in evolution, existing features tend to be
more-or-less ``frozen'' because they are intertwined with the rest of
the organism while it is the new features which are mostly optimized
in relation to the existing frozen features.

Of course, it is quite possible that, at a later stage, a frozen
feature could soften again; especially feedback from the environment
or continuous time might bring about such relaxation and a subsequent
co-evolution of clocks. However, the analysis of this case becomes
significantly more intricate and we will revisit this in the future.

\begin{figure}[htb]
\centering
\input{figs/robu.frozen-bag-info-curve.tex}
\caption{Amount of information as the size of the collection grows for a time interval of length $T_{\text{max}} = 5$.}
\label{fig:frozen-bag-info-curve}
\end{figure}
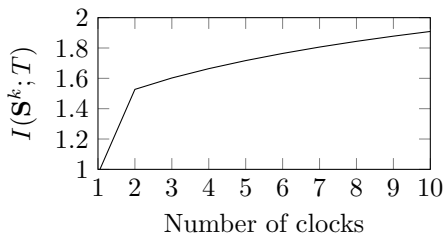

We will now state our model more precisely. We assume that while a
clock is being considered, some $n$ clocks are already in the
collection. Given the state of the clock collection
$\mathbf{S}^n = (S_1,\dots,S_n)$, with $n=0$ indicating the empty
collection, the dynamics of clock $n$ is optimized as to maximize
$I(\mathbf{S}^n;T)$, always for a fixed total duration of
$T_{\text{max}}$; the dynamic parameters of all clocks $k=1,\dots,n-1$
are kept fixed during the optimization. Once the optimization is
complete, clock $n$ is frozen and a new clock $n+1$ is added and the
procedure repeated. In our experiments, we stopped the process once we
reached 10 clocks. The parameter space of this optimization is the set
of probabilities of the transitions $u \rightarrow u$, $u \rightarrow d$,
$d \rightarrow u$ and $d \rightarrow d$ for each of the clocks. The clocks that the
optimization finds
turn out to be either a pure oscillator or else, strictly drop clocks.
As the collection grows, so does the achieved time information
${I(\mathbf{S}^n;T)}$ (Fig.~\ref{fig:frozen-bag-info-curve}). The
first clock added to the bag is the oscillator,
as intuitively expected, as it resolves perfectly 1 full bit of
information. All subsequent additions to the bag, however, turn out to
be pure drop clocks, with no oscillatory component. The first two clocks together add ${\sim 1.5}$ bits, while
every subsequent clock adds significantly less.

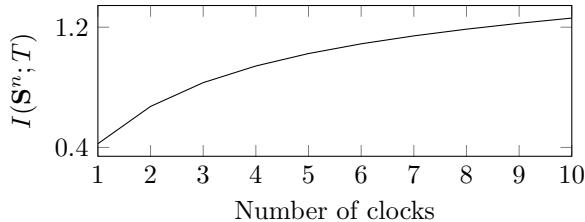
\begin{figure}[htb]
\centering
\input{figs/robu.poisson-info-curve.tex}
\caption{Time information as more identical clocks are added to the set for $T_{\text{max}} = 50$.}
\label{fig:poisson-info-curve}
\end{figure}

Analysing the resulting bag of clocks in detail, we notice that all
the clocks --- except for the first two --- strikingly have very similar dynamics
(parameters are given in Appendix~\ref{sec:darwin-params}). In other
words, once the first two clocks are added, all further clocks
essentially act together as an increasingly refined binomial process,
as more clocks are added\footnote{Special thanks to Nicola
  Catenacci-Volpi for this observation.}.  Because the last clocks
added to the bag have almost identical parameters, we ran another
calculation to explore the behavior of populations of precisely
identical drop clocks. We use a bag of pure drop clocks, all with the
same $r = 0.1$, starting with one of these clocks and adding more
until we reach a total of 10 clocks. We compute $I(\mathbf{S}^n;T)$
for a time scale of size $50$ to create
Figure~\ref{fig:poisson-info-curve}. The curve in this plot shows same diminishing returns from increasing the size of a clock bag, similar to the curve in the previous plot (Fig.~\ref{fig:frozen-bag-info-curve} beyond $N = 2$).

\section{The Clock Cascade}

The clock bag above contained independent clocks which were not
permitted to interact with each other. With the clock cascade,
however, we introduce a simple interaction scheme. 

We arrange the clocks in a queue and begin the simulation by waiting
for the first clock to drop. Once the first clock has dropped, the
second clock is released and may drop (as all drop clocks do,
probabilistically) when time advances after being triggered. The fall
of the second clock, in turn, triggers the release of the third clock
and so on, creating a ``domino'' effect. Finally, after the sequence
has run through all of the clocks, they lie dormant.

Of course, we could just consider a single decaying counter instead,
but we would like here to emphasize that the cascaded counter is
constructed from simple individual clocks and the cascade constitutes
a biologically relevant architecture \citep{nutsch2003signal}.

\subsection{An Illustrative Example}

We take the example of an arrangement of $N = 6$ clocks. We choose this small number here so that the conditional probability (between the state of the arrangement and time) can be plotted easily.

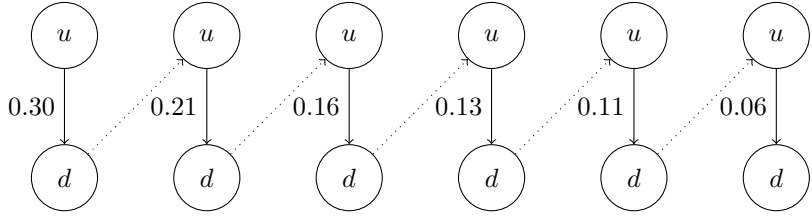
\begin{figure}[htb]
    \centering
    \input{figs/robu.n6-cascade-states.tex}
    \caption{The optimal clock cascade for $N = 6$ and $|T| = 100$. The arrangement contains 6 clocks and they are all in state \emph{up} at the beginning of the simulation. Then, after some time, the first clock drops and by dropping it releases the second clock (notice the dotted arrows). One by one, the sequence runs through all the clocks until the last one decays. The numbers shown in the figure are the (rounded) optimal decay probabilities per tick as found through numerical optimization.}
    \label{fig:cascade-example-params}
\end{figure}

We first used the DIRECT Lipschitzian method for
global optimization \citep{jones93:_lipsc_optim_lipsc_const}
on the model to maximize $I(\mathbf{S}^n;T)$, the mutual information between the state of the sequence and time. The solution to this optimization; the optimal sequence of decay parameters for these circumstances, is shown in Figure~\ref{fig:cascade-example-params}.

Although we have described the behavior in time of this clock arrangement, we have not described how it could be used as a clock. We do so now. As stated previously, a device is a clock if it correlates to time. But, simply knowing that this correlation exists is not sufficient for an agent that wishes to measure time: it is also necessary to know how the states correspond to time. 

\begin{figure}[htb]
    \centering
    \subfloat[\emph{Plot Not stacked}. The dominant region in time of each clock can be seen more clearly in this plot.]{{
        \input{figs/robu.n6-cascade-lines.tex}
    }}
    \quad
    \subfloat[\emph{Probabilties stacked}. The gradual shift in time of the probability mass through the sequence of states can be seen in this plot.]{{
        \input{figs/robu.n6-cascade-stacked.tex}
    }}
    \caption{The conditional probability distribution $P(\mathbf{S}^n|T=t)$ of the number of clocks that are still up in time, plotted in two different ways (stacked lines and not stacked).}
    \label{fig:cascade-example-in-time}
\end{figure}
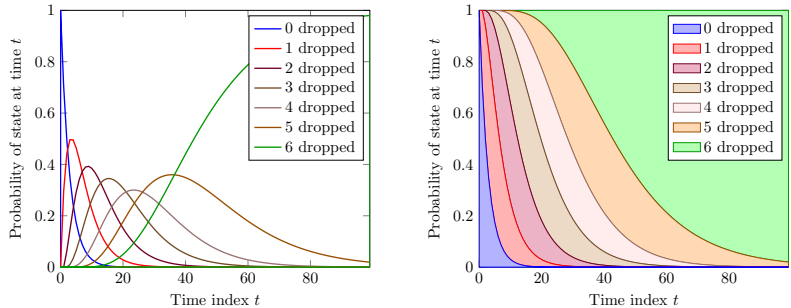

An agent would learn (through experience or evolution or in our case, calculation) what this correspondence is. In the case of the this example, the correspondence produces Figure~\ref{fig:cascade-example-in-time}.

\subsection{The Performance of the Cascade in Different Circumstances}

\begin{figure}[htb]
    \centering
    \input{figs/robu.cascade-grid.tex}
    \caption{The amount of information that the optimal clock cascade has about time for different time windows and different numbers of clocks. We removed the data from the top left corner of the figure and explain the reason in text.}
    \label{fig:clock-cascade-mi-table}
\end{figure}
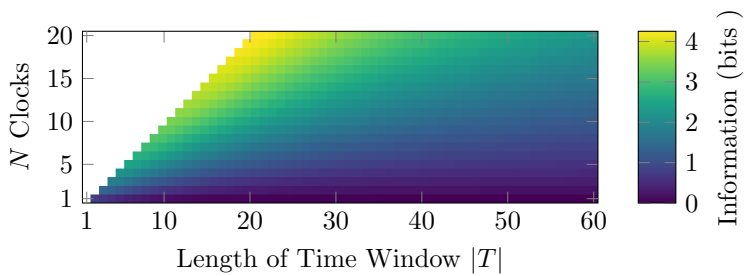

We begin our investigations with this model by computing the maximum performance a clock cascade can attain for different circumstances. Namely, we create cascades having different numbers of clocks $N$ and optimize
those arrangements for different time windows $|T|$. Taken together, these amounts create the plot in Figure~\ref{fig:clock-cascade-mi-table}.

From the plot, we removed the top left corner (where $N \geq |T|$). We did this because $N$ clocks are already sufficient to optimally track a length of time with $N$ ticks and therefore it is not necessary to study longer cascades. The second reason is that the numerical methods that computed this table returned poor (obviously suboptimal) results when the model was configured with more clocks than time ticks. We hypothesize that the Lipschitz algorithm performs poorly when it is given many unused parameters as is the case here in the white region on the plot.

% \laterref{writing in the figure overwrites graphics}

We observe in Fig.~\ref{fig:clock-cascade-mi-table} that the
performance of the clock cascade varies smoothly when more clocks are
added ($N$ is changed) or when the time window $T$ is increased.

\subsection{Clock ``Condensation''}

\begin{figure}[htb]
    \centering
    \input{figs/robu.cascade-condensation.tex}
    \caption{The result of optimizing a cascade of 20 clocks to a time
      window of 100 ticks. The first 8 clocks in the sequence
      ``condense'' (are active only in precisely one time moment each)
      while the rest are spread along their respective time window.
      The parameters of each clock are given in Appendix~\ref{sec:cascade-data}.}
    \label{fig:cascade-condensation-in-time}
\end{figure}
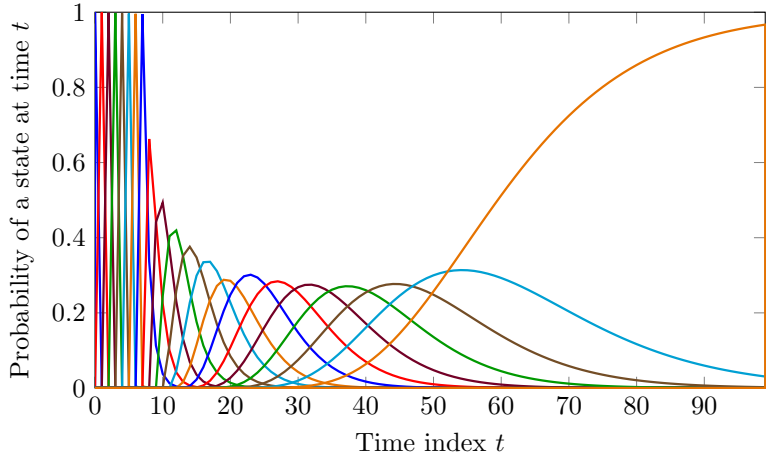

Although Figure~\ref{fig:clock-cascade-mi-table} does not reveal
different regimes (it shows no abrupt changes in clock performance),
we do find for some $N$ and $|T|$ what we consider to be qualitatively
different clock cascades. Specifically, we find a ``condensation''
effect, for example when a cascade of $N = 20$ clocks is optimized for
an axis of time of $|T| = 100$ moments (99 time ticks). The way this clock cascade
evolves in time is plotted in
Figure~\ref{fig:cascade-condensation-in-time}.

To confirm that the condensed clock cascade is a global maximum (and
not an artifact of our choice of optimizer) we also run the same
optimization with a different algorithm to obtain a very similar
result (\ref{sec:psopy-condensation}).

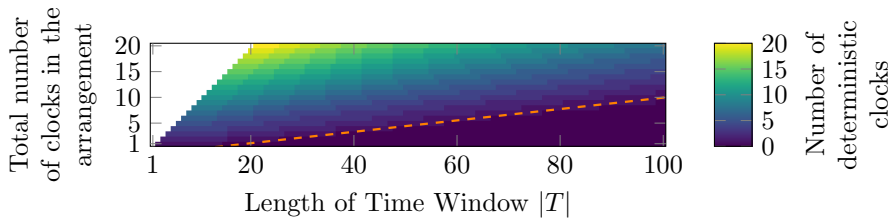
\begin{figure}[htb]
    \centering
    \input{figs/robu.cascade-condensation-grid.tex}
    
    \caption{The amount of condensation for different cascade sizes and different time window sizes. Below the orange dotted line, there is no condensation effect.}
    \label{fig:cascade-condensation-grid}
\end{figure}

To see where this effect occurs, we show a grid plot of the
number of deterministic clocks (that is, the amount of condensation) in the cascade for different $N$ (number of clocks in the cascade in total) and $|T|$ in Figure~\ref{fig:cascade-condensation-grid}. Mainly the plot shows that the condensation gradually happens as clocks are added until the number of clocks matches the number of ticks and then all clocks in the cascade are deterministic.

% \laterref{the writing  in the figure writes across the graphics; check
% also appendix}

\subsection{Special Moments in Time}

Above, in our calculations with the clock cascade, we have been
maximizing for any information about time that the devices can have
across the whole time interval of interest. In general, this will be
quite an abstract quantity. We expect that a more typical necessity
for an organism will be the ability to predict the arrival of a
specific moment in time.

\begin{figure}[htb]
    \centering
    \subfloat[The behavior of the clock cascade for $N = 6$.]{{
        \includegraphics[width=0.28\textwidth]{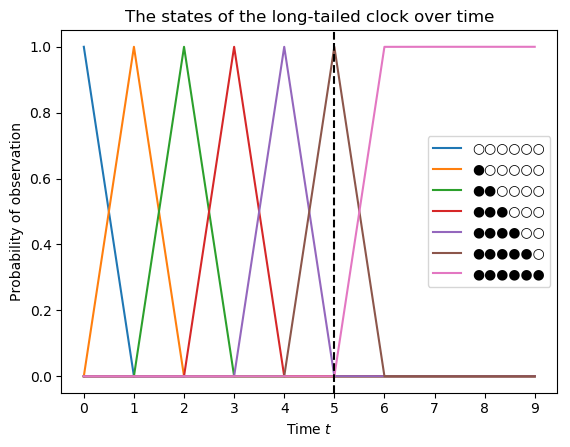}
    }}
    \quad
    \subfloat[And for $N = 5$.]{{
        \includegraphics[width=0.28\textwidth]{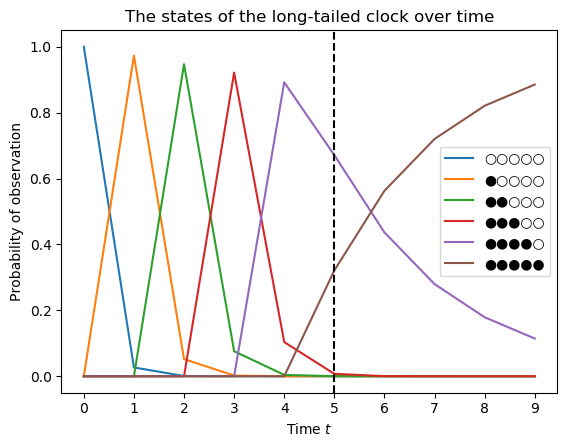}
    }}
    \quad
    \subfloat[The amount of information as clocks are added to the sequence.]{{
        \includegraphics[width=0.28\textwidth]{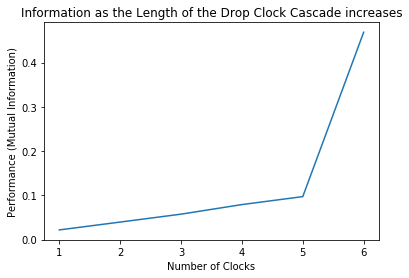}
    }}

    \caption{The optimal clock cascade for when maximizing for information about a particular moment in time.}
    \label{fig:cascade-special-n5}
\end{figure}

We therefore now maximize the amount of information that a clock
cascade\footnote{One reason why we choose to predict a specific moment in
time with the cascade model is that it is well suited
to this task. It gives more information than a bag of drop clocks
(we verified this numerically for all $1 \leq T \leq 6$ and $1 \leq N \leq 5$).
Still, besides the ability to predict time well, clock models can be relevant for
other reasons, such as their robustness or ease of implementation. We leave
the comparison of the robustness or cost of implementation of these models for future work.
} has about a specific moment in time (which we here arbitrarily
picked to be $t = 5$). We repeat the computation for $N = 1, N = 2
\ldots N = 6$ and plot all the clock cascades in this range in
Appendix~\ref{sec:all-six-special-cascades} and we also include the
plot for $N = 5$ here in Figure~\ref{fig:cascade-special-n5}. 

A cascade of six clocks is (of course) able to predict very well the
sixth time moment (marked on the plot with index $t = 5$) as the
arrangement can be made to run deterministically through each clock in
the sequence (Figure~\ref{fig:cascade-special-n5}a). But a cascade of
five clocks is still able to give some\footnote{ Note that in this
  paper we optimize for observation of clock \emph{states} and not
  clock \emph{transitions}. In particular, the observation of the
  clock is assumed to be made by the agent at some unknown random
  moment in time. Therefore, a cascade having less than six clocks
  ($N < 6$) cannot predict $t = 5$ with perfect certainty. Instead,
  one could imagine a different setup in which the agent monitors
  (e.g.\ by continuous polling) the clock and then acts when the last
  clock drops; or, alternatively, one could consider an event-based
  detection of the relevant moment. This alternative setup could
  predict $t = 5$ perfectly with only $N = 5$ clocks (or even $N = 4$
  clocks, depending on how observations and actions are causally
  linked in the model) because it would make use of the mutual
  information between the \emph{transitions} (rather than the states)
  of the cascade and time. Strictly spoken, transitions form ordered
  pairs of states, so being able to identify them would assume a
  hidden or implicit memory in the agent, contrary to our intention to
  model an agent without memory outside of the clock state itself.
  From a formal perspective, using transitions to measure time
  considers a different state space, namely the space of transitions,
  i.e.\ would be translated to our formalism by considering the dual
  of the transition graph. While in a discrete context, only a minor
  reduction in clock complexity is gained by considering transitions
  rather than states, in continuous time, sharp transition events
  offer a significant advantage, and will require a separate detailed
  study.} information about this particular moment in time by having
probabilistic transitions. Figure~\ref{fig:cascade-special-n5}b shows
how, by being probabilistic, the probability of activation of the
fifth clock is delayed from the fifth time moment (where it would have
occurred, had the cascade been deterministic) to partly cover the
sixth moment ($t = 5$).

For a wider picture, it can be seen in
Figure~\ref{fig:cascade-special-n5}c how adding clocks to a cascade
increases the amount of information it can give about the specific
event in time. It increases only gradually as long as the best
solution is probabilistic, but then jumps suddenly as the final clock
is added which completes the deterministic collection.

\section{The Composite Clock}
\label{sec:composite-clock}
Beyond the simplest clocks, clock bags and cascades listed above, we
consider the next more complex clock, a composite clock consisting of
two simple 1-bit clocks each, which, however, are permitted to
communicate (therefore making this arrangement more complex), but with
a constraint on how much communication is permitted. This constraint
is expressed as a measure of information flow between the composite
clocks. This information flow constraint is how we penalize modular
systems for their complexity. Consistent with the freezing
  procedure in the bag-of-clocks experiment we freeze the first clock
  here as well (further comments on ``freezing'' in Sec.~\ref{sec:bag-of-clocks}). In addition to the biological
  justification, the choice of freezing the first clock also leads to more reliable optimization results.
  Thus, as before, the first (upper) component of the
  composite clock becomes an oscillator.
  This composite structure resembles the hierarchical coarse-graining one finds in the algebraic decomposition theory of semigroups and discrete-event dynamical systems \citep{rhodes2009applications,NehanivSYDE710}.

Only the second (lower) component's dynamics will be parametrized and
the parameters optimized according to suitable informational criteria.
This one-way communication is inspired by the semigroup decomposition
of counters \citep{rhodes2009applications}. Future work will allow the
first clock to be optimized as well and add a feedback channel.

We hypothesize that a more comprehensive optimization of the whole
clock system would begin to
optimize the first level of clocks (getting the basic oscillatory
tick), which, when converged, will permit the next level to arise. As the
oscillation is necessarily  optimal for the first level, without
further constraints we do not expect any changes at that level, even
by further joint evolution.

A more interesting scenario arises when the clock ticks become
``soft'' (and there might be additional effects once we permit
feedback between the levels, which is something we will study in the
future). The freezing of a substrate, based on which further, more
intricate patterns can evolve, might be an evolutionarily plausible
mechanism in itself\footnote{Alternatively, depending on the
  configuration, coevolution might also be an option, but for the very
  fundamental set of studies in the present paper, we will not
  consider this further.}. For the present paper, this is the basic
assumption under which we operate: how can, assuming that the first
layer is an optimally converged 1-bit oscillator clock, and assuming
its evolution gets henceforth frozen, the following layer evolve to
make the total clock as effective as possible (with or without
communication).

\begin{figure}[htb]
    \centering
\input{figs/robu.unrolled-composite-clock.tex}
    \caption{The structure of the composite clock unrolled in time. There is a hierarchy of information flow here. Both clocks send information to themselves but only the upper clock sends information to the lower clock.}

\end{figure}
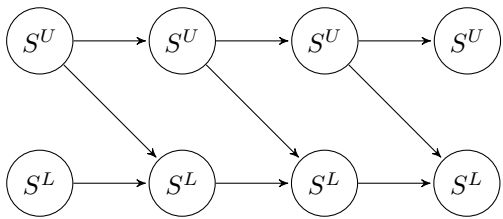
To prepare the description of the full system, we write out in detail
the first component of the clock (the upper component)
as a Markov chain; it is  defined by the matrix $A_U = \left[\begin{smallmatrix}0 & 1 \\ 1 & 0\end{smallmatrix}\right]$ and initial state distribution\footnote{We use the random variable name as a proxy notation for the whole distribution.} $S^U_0 = \left[\begin{smallmatrix}1 \\ 0\end{smallmatrix}\right]$. In the next step, we design the dynamics of the lower component $S^L$, but it is not independent anymore (the stochastic kernel of the lower component is not a square matrix). Rather (with a rectangular matrix as a kernel), it can be influenced by the state of the upper component $U$. The matrix that will drive the behavior of the lower component is a conditional probability distribution, a mapping from the joint state of both components $(S^U_t, S^L_t)$ at time $t$ to the future state of $S^L_{t+1}$ at time $t + 1$. There are 4 columns in this matrix because there are four combinations of the input (i.e.\ condition) states from both components:
$uu$, $ud$, $du$, $dd$ (both up, one up and the other down, etc.).
In vector notation, the complete probability of the whole system to be in a state is written as:
\begin{equation} \label{eq:combine-states}
  P(S^U,S^L) = 
  \begin{bmatrix}
    P(S^U=u,S^L=u) \\
    P(S^U=u,S^L=d) \\
    P(S^U=d,S^L=u) \\
    P(S^U=d,S^L=d)
  \end{bmatrix},
\end{equation}
and the transition matrix representing $P(S^L|S^U,S^L)$ for the lower
clock (we remind that the upper clock can be directly modeled as an
oscillator and that the first coordinate of $S^L$ is the probability
for $u$ and the second the probability for $d$) as the following.

\begin{equation}
A_L =
\left[\begin{smallmatrix}\theta_1 & \theta_2 & \theta_3 & \theta_4 \\
    1 - \theta_1 & 1 - \theta_2 & 1 - \theta_3 & 1 -
    \theta_4\end{smallmatrix}\right].
\label{eq:lower-clock-kernel}
\end{equation}
    
We make use of the fact that
the resulting probabilities add up to 1 and we thus need only one
parameter to describe each of the conditionals.

Combine now the matrix for the upper component ($A_U$) and the matrix
for the lower component ($A_L$) to obtain the complete Markov matrix
for the whole clock:

\[ 
A = 
 \left[\begin{matrix}0 & 0 & \theta_{3} & \theta_{4}\\0 & 0 & 1 - \theta_{3} & 1 - \theta_{4}\\\theta_{1} & \theta_{2} & 0 & 0\\ 1 - \theta_{1} & 1 - \theta_{2} & 0 & 0\end{matrix}\right].
\]

We initialize the lower clock $S^L$, as always, in state $u$, i.e.\
with probability $ \left[\begin{smallmatrix}1 \\
    0\end{smallmatrix}\right]$. Now all required definitions are
complete. Expressed in terms of the joint variable $S= (S^U,S^L)$, and
following the conventions from (\ref{eq:combine-states}), the initial
state of the complete clock is
$S_0 = \left[\begin{smallmatrix}1 & 0& 0& 0\end{smallmatrix}\right]'$
(where $'$ is transpose). Using $A$, the matrix that ticks the clock
forward to its next time step, we can simulate the clock starting at
time 0 and into a future time $t$ by repeated matrix multiplication:
$A^t$. For illustration, the first few states look like:

\begin{figure}[htb]
    \centering
\input{figs/robu.table-sandwich-in-time.tex}
    \caption{The probabilistic state of the composite clock for $t = 0$, $t = 1$ and $t = 2$.}
\end{figure}

\subsection{Experiments with the Composite Clock}
\label{sec:composite-clock-experiment}
The current experiments distinguish themselves from the earlier ones
by having the participating clocks communicate: we create an
information ``tap'' between the two clocks. Importantly, we will
control and limit the amount of information that may flow from the
upper to the lower clock. As discussed earlier, all costs/rewards
(e.g.\ flow vs.\ time resolution) will be expressed exclusively in
terms of Shannon information as unique currency in which the quality
of time measurement is expressed. The expectation is that as more information
is allowed to pass, the overall performance of the composite clock would increase
as the components would be more coordinated.

In the same spirit as in the bag-of-clocks experiment, we assume a
greedy pre-optimization of the upper clock, which
results in an oscillator whose parameters will be frozen. Thus, the
experiment will only optimize the remaining parameters of the clock,
i.e.\ the dynamics of the lower clock and the parameters of its
dependence on the upper clock (in other words, the parameters $\theta_1$,
$\theta_2$, $\theta_3$ and $\theta_4$ of Eq.~\ref{eq:lower-clock-kernel}).
While we maximize time information as before, a
constraint $C$ is imposed on the capacity of communication
\citep[\emph{transfer entropy,}][]{Schreiber2000} from the upper clock
to the lower one ($T$ in the index denotes marginalization over all
valid times occurring in the experiment). Since the system is
Markovian, the flow only over a single step is considered. Since the
upper clock is fixed as the oscillator, no feedback channel is
included. We now compute
\begin{equation*}
 \max_{I(S^U_T;S^L_{T+1}|S^L_T) \leq C} I(S;T)\;.  
\end{equation*}

\begin{figure}[htb]
\centering
\input{figs/robu.dragged_counter_tradeoff.tex}
\caption{In red, the optimum curve found by the DIRECT global Lipschitz optimizer. In black, the more complete curve found by the COBYLA local optimizer. Diagrams of the clocks at the points of interest $a$, $b$ and $c$ are given in Figure~\ref{fig:salient-clocks}. This plot has a minor error which we discovered after performing the numerical optimizations: see Appendix~\ref{sec:sandwich-fix} for the comparison.}
\label{fig:time-information-coupled-clocks}
\end{figure}
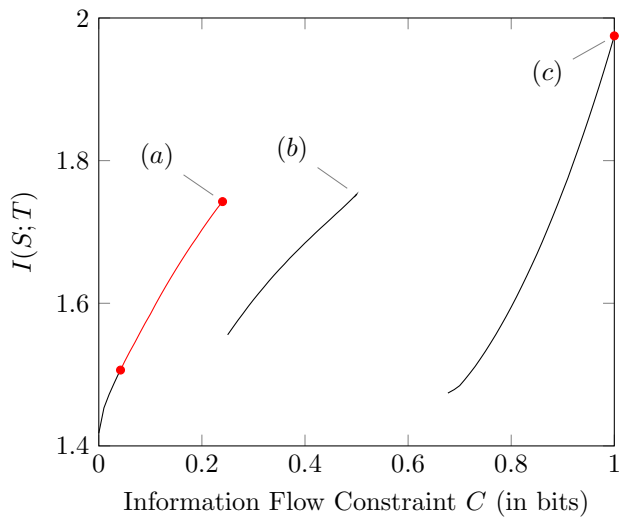

To optimize under the constraint, we use the Lagrangian method,  i.e.\
we maximize $I(S;T) - \lambda I(S^U_T;S^L_{T+1}|S^L_T)$ with Lagrange
parameter $\lambda$.  Scanning through the possible values of
$\lambda$, we cover the spectrum of clocks that arise throughout all
possible constraints. We first used the Lipschitz algorithm to find the red part of the curve in Figure~\ref{fig:time-information-coupled-clocks}. One finds two distinct regimes: the perfect clock at $C=1$, and distinctly suboptimal clocks in the regime below around $C\approx0.2$. The curve breaks off at the low end at about $C=0.04$ due to memory limitations of the Lipschitzian optimizer.

Since the landscape is very complex, in addition to DIRECT, we also applied COBYLA \citep{powell94} to map additional regions of the Lagrangian landscape and to attain an as complete overview as possible over the solution space. With COBYLA, we used
a relaxation method to cover locally optimal parts of the curve
which the global Lipschitz optimizer does not access. Concretely, we
started optimization at the most permissive information flow bound
$C=1$, optimized, and then tightened the bound slowly, always starting
with the clock parameters obtained for the previous constraint $C$.

The results are shown again in
Fig.~\ref{fig:time-information-coupled-clocks}, but they now include
the black regions in addition to the red. Since this optimization is
not global, it is able to uncover additional structure in the solution
space. First of all, when, starting at the optimal clock $C=1$, $C$ is
reduced, the trade-off curve between the constraint $C$ and the time
information $I(S;T)$ falls below the globally
optimal solutions. The ensuing solutions correspond to ``fuzzy'' counters (the optimal clock is a binary counter), but do not trade in information flow and achieved time information in a well-balanced way, although part of that portion of the curve is still Pareto-optimal (is not superseded simultaneously by solutions better in terms of both $C$ \emph{and} $I(S;T)$). 

In the range $C\approx0.5-0.7$, no solution is found. Between
$C\approx0.25-0.5$, a class of locally optimal solutions is found
which is not Pareto optimal.
Finally, below $C=0.25$, one regains the lower $C$ regime which is
found with global optimization. The curve continues down to $C=0$
(continuation of the red to the black curve), which we obtain with
COBYLA which is not suffering from the memory problems of the global
optimizer. The two clock classes below $C\approx0.5$ look very similar
as a whole clock, but distribute the ``counting'' differently over
their component clocks.

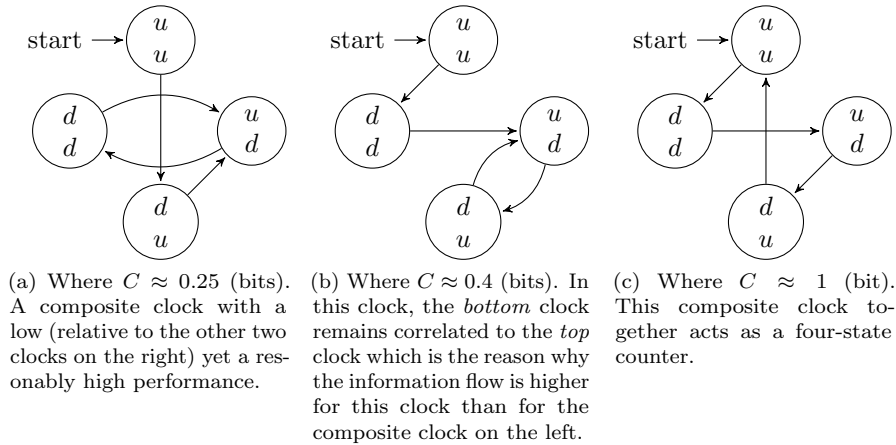
\begin{figure}[htb]
  \centering
  \subfloat[Where $C \approx 0.25$ (bits). A composite clock with a low (relative to the other two clocks on the right) yet a resonably high performance. 
  ]{{
    \input{figs/robu.salient-sandwich-0.25.tex}\unskip
  }}
  \quad
  \subfloat[Where $C \approx 0.4$ (bits). In this clock, the \emph{bottom} clock remains correlated to the \emph{top} clock which is the reason why the information flow is higher for this clock than for the composite clock on the left.]{{
    \input{figs/robu.salient-sandwich-0.4.tex}\unskip
  }}
  \quad
  \subfloat[Where $C \approx 1$ (bit). This composite clock together acts as a four-state counter.]{{
    \input{figs/robu.salient-sandwich-1.tex}\unskip
  }}%
  \caption{Above, the composite clocks of salient points in the tradeoff in Fig.~\ref{fig:time-information-coupled-clocks}. The diagrams show transitions of the composite clock in pairs of states of both the \emph{top} and the \emph{bottom} clock.}%
  \label{fig:salient-clocks}
\end{figure}

Apart from the discovery of different clock ``regimes'', a detailed
inspection of the space of possible configurations (Fig.~\ref{fig:salient-clocks})
demonstrates that, in part, finding parameters which achieve high
values of time information is very difficult. Furthermore, despite
large permitted flows $C$, time resolution may still have low values.
This makes it clear that measuring
time is not an incidental effect that is likely to be found en passant
by an evolutionary process; instead, one rather will expect good time
measurement abilities to have been explicitly evolved for, either
directly or indirectly (via proxy criteria). In a hierarchy any more
complex than the one discussed here, we expect the clocks to need to
be attuned to each other for optimal resolution.

\section{Final Comments and Future Work}

We have studied the ability of minimal clocks to resolve time
information. In particular, even 1-bit clocks can, as drop clocks,
provide information about global time if the overall time horizon is
known when the clock parameters are set. 

Let us summarize one of the central insights: the only features of our
axis of time are its total maximal extent and its grain. The
oscillator matches the grain and represents local relative time
information and the drop clock matches the overall duration, i.e.\ the
global information about the current point in time. Since the clock is
Markovian it is per its nature time-local; it is remarkable that it is
possible to attune the drop probability to the global time period of
interest to effectively extract global time information with a
limited, purely local mechanism. In evolution, such a probability can
be expected to evolve over many generations for the optimal resolution
of the time intervals of interest.

The relation of measured interval and clock parameters is not
straightforward and is marked by an interplay of global and local
properties (Fig.~\ref{fig:two-maxima-drop-clock}). When extending to a
bag-of-clocks, the first two clocks are an oscillator and a drop clock
of a particular time constant, followed by further drop clocks, which
are then, however, nearly identical; thus, apart from the first two,
the rest of the clocks operate as a nearly binomial process. A cascade of clocks
can stretch to reach target moments in time or condense in narrow time windows.
Finally,
when two clocks are stacked together with limited communication, a
rich set of regimes opens up, of which just two, the perfect clock,
and a very soft clock, can be Lagrange-optimal. These studies provide
a spectrum of candidates for behaviors of minimal clocks which one
could try to identify in biological systems.

One central limitation of the present work is the assumption of a
fixed, global tick which drives all the clocks. Future work will,
therefore, include the consideration of clocks in continuous time,
where the dynamics needs to establish and sustain a synchronization
between the subclocks in addition to the coordination of their
respective resolution regime. Additionally, we conjecture that
informationally optimal clocks will exhibit some robustness for the
resolution of relevant time regimes. This will form the basis for
future studies.

\section{Acknowledgements}
We thank Nicola Catenacci-Volpi, Simon Smith and Adeline Chanseau for
discussions on this topic, as well as the anonymous reviewers of an
earlier version of the paper for very useful
questions and comments.

Christoph Salge is funded by the EU Horizon 2020 program under the
Marie Sklodowska-Curie grant 705643.

\footnotesize
\bibliographystyle{apa-good}
%\bibliography{vitpub,references,Alife}
\bibliography{robu}

\begin{appendices}
    \section{Drop Clock Bag}
    \subsection{Parameters of Incremental Optimization}
    \label{sec:darwin-params}

    Results of optimizing a bag of 10 clocks with incremental freezing. The DIRECT Lipschitz optimizer was used set to 30 iterations.

    \begin{center}
        \begin{tabular}{cll} \hline
        Clock \# & Probability $u \to d$ & Probability $d \to u$ \\ \hline
        1 & 0.0000084675439042 & 0.9999915324560957 \\
        2 & 0.0727277345933039 & 0.0000254026317127 \\
        3 & 0.4945130315500686 & 0.0002286236854138 \\
        4 & 0.5187471422039323 & 0.0002286236854138 \\
        5 & 0.5187471422039323 & 0.0002286236854138 \\
        6 & 0.5187471422039323 & 0.0002286236854138 \\
        7 & 0.5192043895747599 & 0.0002286236854138 \\
        8 & 0.5205761316872427 & 0.0006858710562415 \\
        9 & 0.5205761316872427 & 0.0006858710562415 \\
        10 & 0.5205761316872427 & 0.0006858710562415 \\ \hline
        \end{tabular}
    \end{center}

    \section{Drop Clock Cascade}
    \subsection{Parameters of a Condensed Cascade}
    \label{sec:cascade-data}

    Here we show the probability to drop of each step in a cascade of 20
    clocks as found by the DIRECT Lipschitz optimizer after 1000
    iterations when maximizing for the overall information about time
    (not specifically about a particular moment). The table shows that the parameters
    for the first seven clocks are close to one.

    \begin{center}
        \begin{tabular}{cl} \hline
        Clock \# & Drop Probability \\ \hline
        1 & 0.9993141289437585 \\
        2 & 0.9993141289437585 \\
        3 & 0.9993141289437585 \\
        4 & 0.9993141289437585 \\
        5 & 0.9993141289437585 \\
        6 & 0.9993141289437585 \\
        7 & 0.9993141289437585 \\
        8 & 0.6659807956104252 \\
        9 & 0.6659807956104252 \\
        10 & 0.5548696844993141 \\
        11 & 0.4684499314128944 \\
        12 & 0.3984910836762689 \\
        13 & 0.3436213991769547 \\
        14 & 0.3340192043895748 \\
        15 & 0.2572016460905350 \\
        16 & 0.2229080932784636 \\
        17 & 0.1886145404663923 \\
        18 & 0.1570644718792867 \\
        19 & 0.1241426611796982 \\
        20 & 0.0829903978052126 \\ \hline
        \end{tabular}
    \end{center}

    \subsection{Alternative Optimization Also Finds Condensation}
    \label{sec:psopy-condensation}

    Here, we run the same experiment of optimizing a cascade of 20
    clocks to a time window of 100 moments, but instead of using
    DIRECT Lipschitz, we optimize using a particle swarm optimizer.
    Although the results are noisy, they increase our confidence in
    the validity of the ``condensation'' solution found by Lipschitz.
    See plot below (Fig.~\ref{fig:psopy-condensation}).

    \begin{center}
        \includegraphics[width=0.4\textwidth]{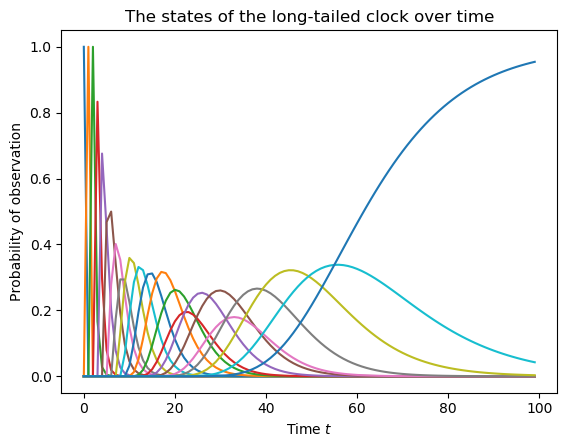}
        \captionof{figure}{Alternative optimization of the clock cascade also finds the condensation effect.}
        \label{fig:psopy-condensation}%
    \end{center}

    \subsection{Identifying Specific Moments in Time}
    \label{sec:all-six-special-cascades}

    The behavior of the clock cascade for $N = 1, N = 2 \ldots N = 6$
    optimized to give information about a specific moment in time at
    $t = 5$:

    \begin{center}
        \includegraphics[width=0.28\textwidth]{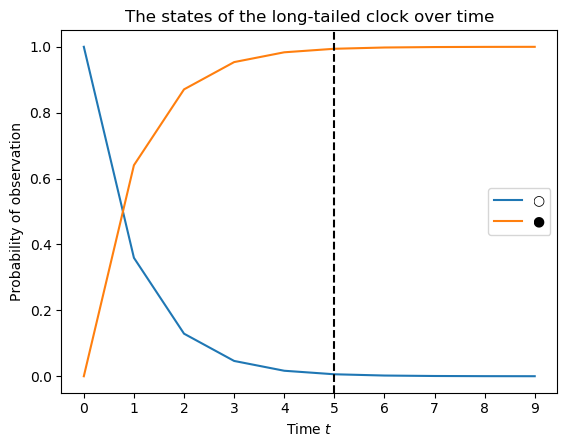}
        \quad
        \includegraphics[width=0.28\textwidth]{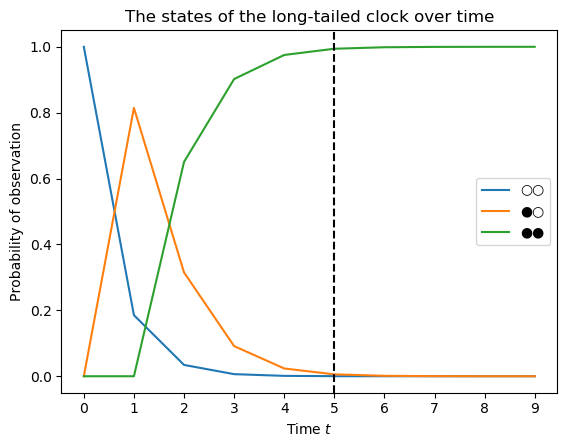}
        \quad
        \includegraphics[width=0.28\textwidth]{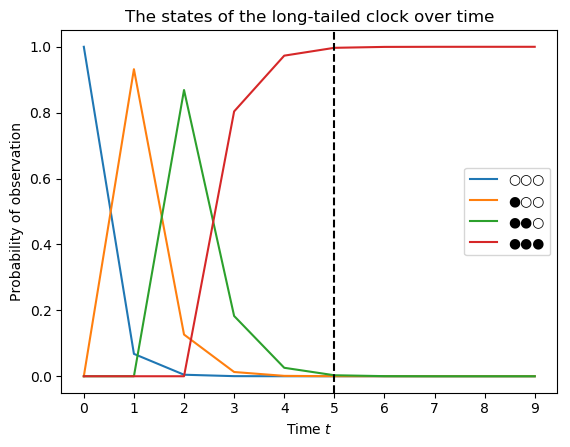}

        \includegraphics[width=0.28\textwidth]{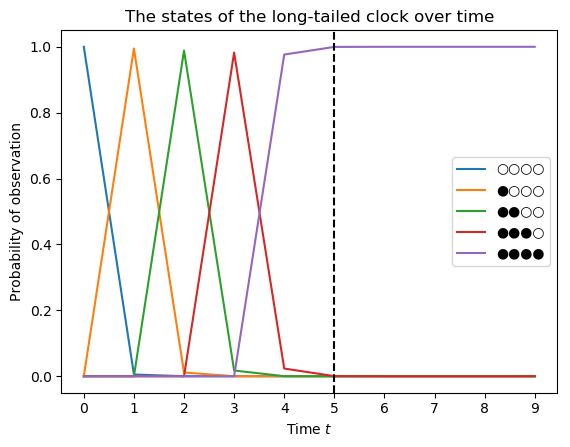}
        \quad
        \includegraphics[width=0.28\textwidth]{figs/robu.clock-cascade-target-n5.png}
        \quad
        \includegraphics[width=0.28\textwidth]{figs/robu.clock-cascade-target-n6.png}
    \end{center}
    \captionof{figure}{The figure shows in detail how optimal clock cascades stall (lengthen the average amount of time until the decay of their final clock) by decaying at every step probabilistically.}

    \section{Composite Clock Computation Error}
    \label{sec:sandwich-fix}

    \begin{center}
      \includegraphics[width=0.4\textwidth]{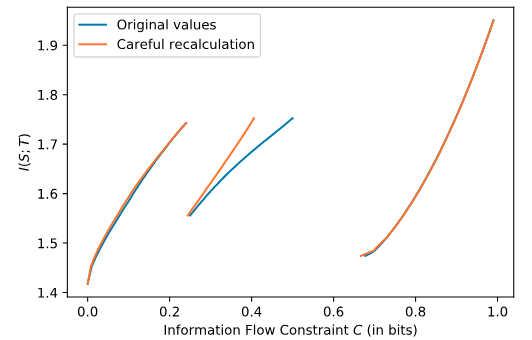}
  \end{center}
  
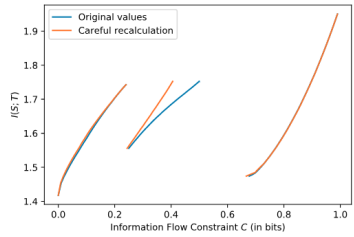
\captionof{figure}{The landscape plotted here is a careful recomputation of Figure~\ref{fig:time-information-coupled-clocks} showing the same qualitative features but fixing minor errors (in the constraint function, specifically) which do not affect the overall character of the solution.}

\end{appendices}
   
\end{document}

%% file: figs/robu.alternator-state-diagram.tex
\begin{tikzpicture}[>=stealth',shorten >=1pt,auto,node distance=2cm]
  \node[initial,state] (u)      {$u$};
  \node[state]         (d) [right of=u]  {$d$};

  \path[->]
      (u) edge [loop above] node {$1 - r$} (u)
      (u) edge [bend left]  node {$r$} (d)
      (d) edge [loop above] node {$1 - r$} (d)
      (d) edge [bend left]  node {$r$} (u);
\end{tikzpicture}

%% file: figs/robu.alternator-time-behaviour.tex
\begin{tikzpicture}
\begin{axis}[
    ymin = 0, ymax = 1,
    xmin = 0, xmax = 20,
    xlabel = $t$,
    ylabel = {$p(u \mid t)$},
]
\addplot [
    domain=0:20, 
    samples=21
]
{0.5*(1+(1-2*0.95)^x)};
\addlegendentry{$r = 0.95$}
\addplot [
    domain=0:20, 
    samples=21, 
    dashdotted
]
{0.5*(1+(1-2*0.05)^x)};
\addlegendentry{$r = 0.05$}
\addplot [
    domain=0:20, 
    samples=21, 
    dashed
]
{0.5*(1+(1-2*0.5)^x)};
\addlegendentry{$r = 0.5$}
\end{axis}
\end{tikzpicture}

%% file: figs/robu.drop-clock-state-diagram.tex
\begin{tikzpicture}[>=stealth',shorten >=1pt,auto,node distance=2cm]
  \node[initial,state] (u)      {$u$};
  \node[state]         (d) [right of=u]  {$d$};

  \path[->]
      (u) edge [loop above] node {$1 - r$} (u)
      (u) edge              node {$r$} (d);
\end{tikzpicture}

%% file: figs/robu.drop-clock-phase-transition.tex
\begin{tikzpicture}
\begin{axis}[
    xlabel=Timespan,
    ylabel=Optimal $r$,
    xmin=2,xmax=100,
    ymin=0,ymax=1.1,
    height=3.6cm,
    width=6cm
]

\addplot [only marks, mark=*, mark size=0.25] table[x index=0, y index=1] 
    {data/robu.bestr_known_start_100_1000.dat};
\end{axis}
\end{tikzpicture}

%% file: figs/robu.drop-clock-2d-plot.tex
\begin{tikzpicture}
\begin{axis}[
  axis on top,
  xmin=1.5,xmax=30.5,ymin=0,ymax=1,
  colorbar,
  colormap/viridis,
  colorbar style={ylabel={Information (bits)}},
  xlabel=Timespan Size,
  ylabel=Drop Probability,
  extra x ticks={2},
  extra x tick labels={$2$},
  x=5cm/31,y=4cm
]
% This png comes from the awesome-drop-clock-p-vs-t-grid-plot.ipynb
% notebook.
\addplot graphics [xmin=1.5,xmax=30.5,ymin=0,ymax=1]
  {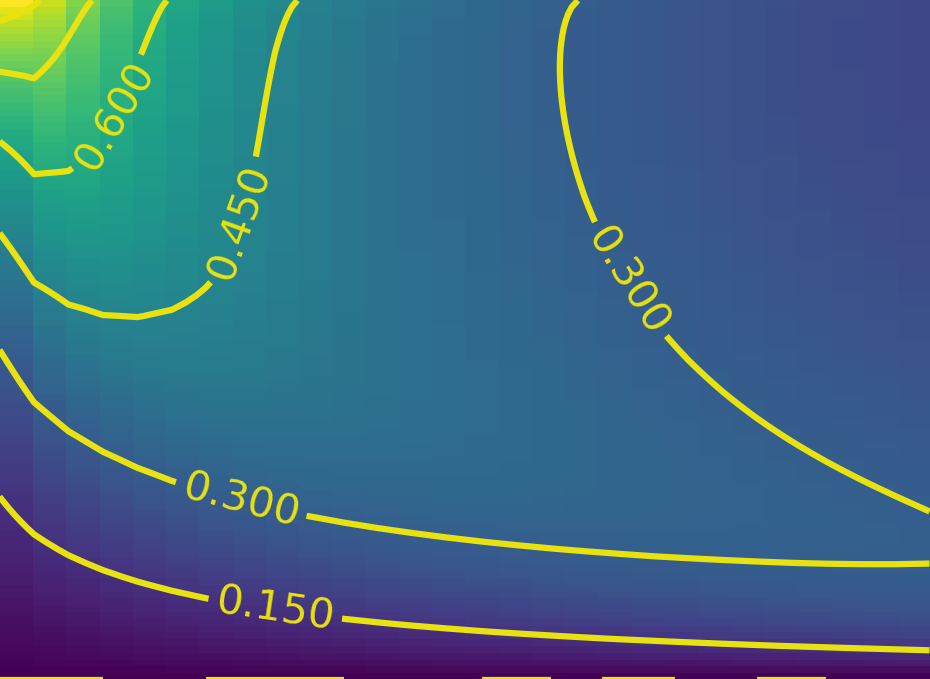};

\addplot [only marks, yellow, mark size=0.8pt] table[x index=0, y index=1] 
{data/robu.bestr_known_start_100_1000.dat};

\draw [dashed, line width=0.3mm, color=orange] (20,0) -- (20,1);
\end{axis}
\end{tikzpicture}

%% file: figs/robu.two-maxima-drop-clock.tex
\begin{tikzpicture}
\begin{axis}[
  width=7cm,
  legend style = { at = {(1.3, 0.65)}},
  legend cell align={left},
  xlabel=Drop Probability $r$,
  ylabel=$I(S;T)$,
  xmin=0,xmax=1,
  ymin=0,ymax=0.4,
]

\addplot [] table[x index=0, y index=1] {data/robu.known_start_reldrop_20.dat};
\addlegendentry{$I(S; T)$};
\addplot [dotted, thick] table[x index=0, y index=2] {data/robu.known_start_reldrop_20.dat};
\addlegendentry{$I(S; \input{figs/robu.lr-variable.tex}\unskip)$};
\addplot [dashed] table[x index=0, y index=3] {data/robu.known_start_reldrop_20.dat};
\addlegendentry{$I(S; \input{figs/robu.alt-variable.tex}\unskip)$};

\end{axis}
\end{tikzpicture}

%% file: figs/robu.lr-variable.tex
\crule[blue]{0.2cm}{0.2cm}\crule[blue]{0.2cm}{0.2cm}\crule[blue]{0.2cm}{0.2cm}\crule[red]{0.2cm}{0.2cm}\crule[red]{0.2cm}{0.2cm}\crule[red]{0.2cm}{0.2cm}

%% file: figs/robu.alt-variable.tex
\crule[blue]{0.2cm}{0.2cm}\crule[red]{0.2cm}{0.2cm}\crule[blue]{0.2cm}{0.2cm}\crule[red]{0.2cm}{0.2cm}\crule[blue]{0.2cm}{0.2cm}\crule[red]{0.2cm}{0.2cm}

%% file: figs/robu.frozen-bag-info-curve.tex
\begin{tikzpicture}
\begin{axis}[
  width=6cm,
  height=3.6cm,
  xlabel=Number of clocks,
  ylabel=$I(\mathbf{S}^k;T)$,
  xmin=1,xmax=10,
  ymin=1,ymax=2,
  xtick distance={1}
]

  \addplot [] table[x index=0, y index=1] 
    {data/robu.darwin_lipschitz_mis_5.dat};
%    \addplot [] table[x index=0, y index=1] {data/robu.parralel_10_60.dat};
%    \addlegendentry{$t_{max} = 5$};

\end{axis}
\end{tikzpicture}

%% file: figs/robu.poisson-info-curve.tex
\begin{tikzpicture}
\begin{axis}[
xlabel=Number of clocks,
ylabel=$I(\mathbf{S}^n;T)$,
xmin=1,xmax=10,
y=2cm, ytick={0.4,1.2},
x=0.7cm, xtick={1, ...,10}
]

\addplot [] table[x index=0, y index=1] 
    {data/robu.parralel_10_5.dat};
%    \addlegendentry{$t_{max} = 60$};
%    \addplot [] table[x index=0, y index=1] {data/robu.parralel_10_60.dat};
%    \addlegendentry{$t_{max} = 5$};

\end{axis}
\end{tikzpicture}

%% file: figs/robu.n6-cascade-states.tex
\begin{tikzpicture}

    \setsepchar{,}
    % These parameters come from
    % the lipschitz/sequence-of-clocks.ipynb
    % notebook.
    \readlist\term{0.30,0.21,0.16,0.13,0.11,0.06}

    \node[state] (u1) {$u$};
    \node[state] (d1) [below=of u1] {$d$};
    \path[->] (u1) edge node[left] {$\term[1]$} (d1);
    \foreach \i [
        evaluate=\i as \prev using int(\i - 1)
    ] in {2,...,6} {
        \node[state] (u\i) [right=of u\prev] {$u$};
        \node[state] (d\i) [below=of u\i] {$d$};
        \path[->] (u\i) edge node[left] {$\term[\i]$} (d\i);
        \path[->, dotted] (d\prev) edge node {} (u\i);
    }
\end{tikzpicture}

%% file: figs/robu.n6-cascade-lines.tex
\begin{tikzpicture}[scale=0.6]
    \begin{axis}[
        xlabel={Time index $t$},
        ylabel={Probability of state at time $t$},
        cycle list={{blue,mark=none},
        {red,mark=none},
        {purple!60!black,mark=none},
        {brown!60!black,mark=none},
        {pink!60!black,mark=none},
        {orange!60!black,mark=none},
        {green!60!black,mark=none}},
        enlargelimits=false
        ]
    \foreach \i [
        evaluate=\i as \dropped using int(\i - 1)
    ] in {1,...,7} {
        \addplot+[line width=0.3mm] table[x index=0, y index=\i] 
        {data/robu.n6-cascade-cond.dat} \closedcycle;
        \addlegendentryexpanded{\dropped{} dropped};
    }
    \end{axis}
\end{tikzpicture}

%% file: figs/robu.n6-cascade-stacked.tex
\begin{tikzpicture}[scale=0.6]
    \begin{axis}[
        xlabel={Time index $t$},
        ylabel={Probability of state at time $t$},
        stack plots=y,
        cycle list={{blue,fill=blue!30!white,mark=none},
        {red,fill=red!30!white,mark=none},
        {purple!60!black,fill=purple!30!white,mark=none},
        {brown!60!black,fill=brown!30!white,mark=none},
        {pink!60!black,fill=pink!30!white,mark=none},
        {orange!60!black,fill=orange!30!white,mark=none},
        {green!60!black,fill=green!30!white,mark=none}},
        area legend,
        enlargelimits=false
        ]
    \foreach \i [
        evaluate=\i as \dropped using int(\i - 1)
    ] in {1,...,7} {
        \addplot table[x index=0, y index=\i] 
        {data/robu.n6-cascade-cond.dat} \closedcycle;
        \addlegendentryexpanded{\dropped{} dropped};
    }
    \end{axis}
\end{tikzpicture}

%% file: figs/robu.cascade-grid.tex
\begin{tikzpicture}
    \begin{axis}[
        enlargelimits=false,axis on top,
        colorbar,
        colormap/viridis,
        colorbar style={ylabel={Information (bits )}},
        point meta min=0, point meta max=4.253532,
        xlabel={Length of Time Window $|T|$},
        ylabel={$N$ Clocks},
        axis equal image,
        extra x ticks={1},
        extra x tick labels={$1$},
        extra y ticks={1},
        extra y tick labels={$1$},
        % x=0.01cm, y=0.01cm
        ]
        \addplot graphics[
          ymin=0.5, ymax=20.5, 
          xmin=0.5, xmax=60.5
          ]
            {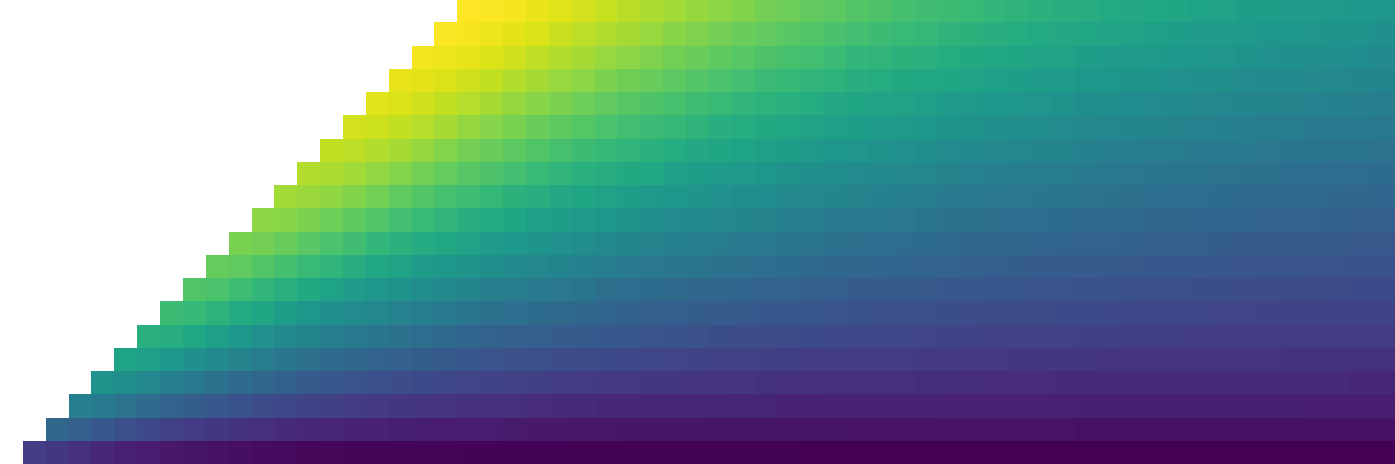};
            % The data was created in lipschitz/sequential-clock-mi-grid.ipynb
            % The data of this figure is available in
            % data/robu.cascade-grid.dat
    \end{axis}
\end{tikzpicture}

%% file: figs/robu.cascade-condensation.tex
\begin{tikzpicture}
    \begin{axis}[
        xlabel={Time index $t$},
        ylabel={Probability of a state at time $t$},
        cycle list={{blue,mark=none},
        {red,mark=none},
        {purple!60!black,mark=none},
        {green!60!black,mark=none},
        {brown!60!black,mark=none},
        {cyan!85!black,mark=none},
        {orange!90!black,mark=none}},
        enlargelimits=false,
        x=9cm/100,
        y=5cm
        ]
    \foreach \i [
        evaluate=\i as \dropped using int(\i - 1)
    ] in {1,...,21} {
        \addplot+[line width=0.3mm] table[x index=0, y index=\i]
        {data/robu.clock-cascade-condensation.dat} \closedcycle;
    }
    \end{axis}
\end{tikzpicture}

%% file: figs/robu.cascade-condensation-grid.tex
\begin{tikzpicture}
    \begin{axis}[
        enlargelimits=false,axis on top,
        colorbar,
        colormap/viridis,
        colorbar style={ylabel style={align=center}, ylabel={Number of \\ deterministic \\ clocks}},
        point meta min=0, point meta max=20,
        xlabel={Length of Time Window $|T|$},
        ylabel style={align=center},
        ylabel={Total number \\ of clocks in the \\ arrangement},
        axis equal image,
        extra x ticks={1},
        extra x tick labels={$1$},
        extra y ticks={1},
        extra y tick labels={$1$},
        % x=0.01cm, y=0.01cm
        ]
        \addplot graphics[
          ymin=0.5, ymax=20.5, 
          xmin=0.5, xmax=100.5
          ]
            {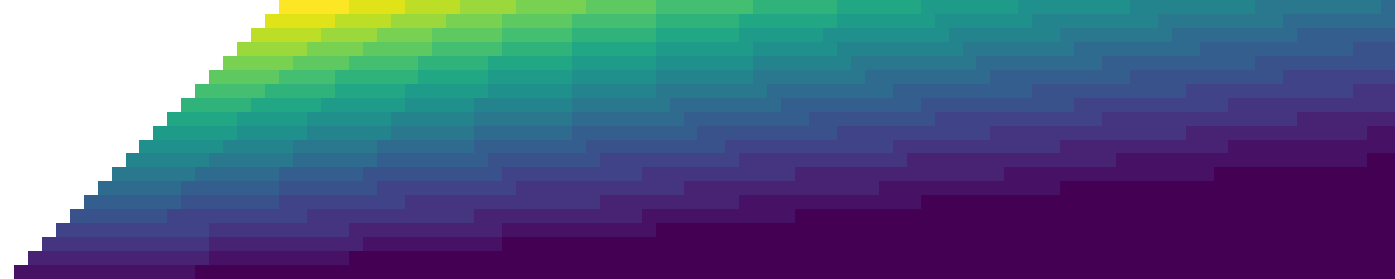};
            % The data was created in lipschitz/sequential-clock-mi-grid.ipynb

        \draw [dashed, line width=0.3mm, color=orange] (10,0) -- (100.5, 10);
    \end{axis}
\end{tikzpicture}

%% file: figs/robu.unrolled-composite-clock.tex
\begin{tikzpicture}[
      >=stealth',
      shorten >=1pt,
      auto
    ]
    \node[state] (u0) {$S^U$};
    \node[state] (l0) [below=of u0] {$S^L$};
    \node[state] (u1) [right=of u0] {$S^U$};
    \node[state] (l1) [right=of l0] {$S^L$};
    \node[state] (u2) [right=of u1] {$S^U$};
    \node[state] (l2) [right=of l1] {$S^L$};
    \node[state] (u3) [right=of u2] {$S^U$};
    \node[state] (l3) [right=of l2] {$S^L$};

    \path[->] (u0) edge node {} (u1);
    \path[->] (u0) edge node {} (l1);
    \path[->] (l0) edge node {} (l1);
    \path[->] (u1) edge node {} (u2);
    \path[->] (u1) edge node {} (l2);
    \path[->] (l1) edge node {} (l2);
    \path[->] (u2) edge node {} (u3);
    \path[->] (u2) edge node {} (l3);
    \path[->] (l2) edge node {} (l3);
\end{tikzpicture}

%% file: figs/robu.table-sandwich-in-time.tex
\def\arraystretch{1.2}
\begin{tabular}{ |p{0.6cm}|p{1.3cm}|p{5cm}|  }
 \hline
 \multicolumn{3}{|c|}{Time dynamics of both parts} \\
 \hline
 $S_0$ & $A S_0$ & $A^2 S_0$ \\
 \hline
 
 &&\\
 $\left[\begin{matrix}1\\0\\0\\0\end{matrix}\right]$ &
 $ \left[\begin{matrix}0\\0\\\theta_{1}\\1 - \theta_{1}\end{matrix}\right]$ &
 $ \left[\begin{matrix}\theta_{1} \theta_{3} + \theta_{4} \left(1 - \theta_{1}\right)\\\theta_{1} \left(1 - \theta_{3}\right) + \left(1 - \theta_{1}\right) \left(1 - \theta_{4}\right)\\0\\0\end{matrix}\right]$ \\
 &&\\

 \hline
\end{tabular}

%% file: figs/robu.dragged_counter_tradeoff.tex
\begin{tikzpicture}
\begin{axis}[
    xlabel=Information Flow Constraint $C$ (in bits),
    ylabel=$I(S;T)$,
    xmin=0, xmax=1,
    ymin=1.4,ymax=2
]

%% \addplot [color=red] table[x index=0, y index=1] 
%%     {data/robu.drag_n9_mi_condmi_a.dat};
\addplot [] table[x index=0, y index=1] 
    {data/robu.drag_n9_mi_condmi_a1.dat};
\addplot [color=red] table[x index=0, y index=1] 
    {data/robu.drag_n9_mi_condmi_a2.dat};
\addplot [] table[x index=0, y index=1] 
    {data/robu.drag_n9_mi_condmi_b.dat};
\addplot [] table[x index=0, y index=1] 
    {data/robu.drag_n9_mi_condmi_c.dat};
\addplot [mark=*, color=red, mark size=1.5] table []
{
0.9999999999999487   1.9749375012017316
};

\addplot[mark=*, color=red, mark size=1.5] coordinates{(0.0421478658,1.5063372572)};
\addplot[mark=*, color=red, mark size=1.5] coordinates{(0.2400902399,1.7425492069)};

\addplot[] coordinates {(0.2400902399,1.7425492069)} node[pin=150:{$(a)$}]{} ;
\addplot[] coordinates {(0.500024987, 1.7519991607)} node[pin=150:{$(b)$}]{} ;
\addplot[] coordinates {(0.99999999, 1.974937)} node[pin=200:{$(c)$}]{} ;

\end{axis}
\end{tikzpicture}

%% file: figs/robu.salient-sandwich-0.25.tex
\begin{tikzpicture}[>=stealth',shorten >=1pt,auto,node distance=1.71cm]
    \node[initial,state,align=center] (uu)      {$u$ \\ $u$};
    \node[state,align=center]         (ud) [below right of=uu]  {$u$ \\ $d$};
    \node[state,align=center]         (du) [below left of=ud]  {$d$ \\ $u$};
    \node[state,align=center]         (dd) [above left of=du]  {$d$ \\ $d$};

    \path[->]
        (uu) edge node {$ $} (du)
        (ud) edge [bend left] node {$ $} (dd)
        (du) edge node {$ $} (ud)
        (dd) edge [bend left] node {$ $} (ud);
\end{tikzpicture}

%% file: figs/robu.salient-sandwich-0.4.tex
\begin{tikzpicture}[>=stealth',shorten >=1pt,auto,node distance=1.71cm]
    \node[initial,state,align=center] (uu)      {$u$ \\ $u$};
    \node[state,align=center]         (ud) [below right of=uu]  {$u$ \\ $d$};
    \node[state,align=center]         (du) [below left of=ud]  {$d$ \\ $u$};
    \node[state,align=center]         (dd) [above left of=du]  {$d$ \\ $d$};
  
    \path[->]
        (uu) edge node {$ $} (dd)
        (ud) edge [bend left] node {$ $} (du)
        (du) edge [bend left] node {$ $} (ud)
        (dd) edge node {$ $} (ud);
  \end{tikzpicture}

%% file: figs/robu.salient-sandwich-1.tex
\begin{tikzpicture}[>=stealth',shorten >=1pt,auto,node distance=1.71cm]
    \node[initial,state,align=center] (uu)      {$u$ \\ $u$};
    \node[state,align=center]         (ud) [below right of=uu]  {$u$ \\ $d$};
    \node[state,align=center]         (du) [below left of=ud]  {$d$ \\ $u$};
    \node[state,align=center]         (dd) [above left of=du]  {$d$ \\ $d$};
  
    \path[->]
        (uu) edge node {$ $} (dd)
        (ud) edge node {$ $} (du)
        (du) edge node {$ $} (uu)
        (dd) edge node {$ $} (ud);
  \end{tikzpicture}